\documentclass[pra,twocolumn,superscriptaddress,longbibliography]{revtex4-2}
\usepackage[T1]{fontenc}
\usepackage[utf8]{inputenc}
\usepackage{amsmath,mathtools,amssymb}
\usepackage{setspace}
\usepackage[colorlinks=true,linkcolor=blue,urlcolor=blue,citecolor=blue]{hyperref}
\usepackage{listings}
\usepackage[dvipsnames]{xcolor}

\DeclareFixedFont{\ttb}{T1}{txtt}{bx}{n}{9} 
\DeclareFixedFont{\ttm}{T1}{txtt}{m}{n}{9} 
\definecolor{deepblue}{rgb}{0,0,0.5}
\definecolor{deepred}{rgb}{0.6,0,0}
\definecolor{deepgreen}{rgb}{0,0.5,0}
\definecolor{orange}{rgb}{1,.498, 0}

\definecolor{bluekeywords}{rgb}{0.13, 0.13, 1}
\definecolor{greencomments}{rgb}{0, 0.5, 0}
\definecolor{redstrings}{rgb}{0.9, 0, 0}
\definecolor{graynumbers}{rgb}{0.5, 0.5, 0.5}

\begin{document}
\setstretch{1}
\title{\texttt{MFDFA}: Efficient Multifractal Detrended Fluctuation Analysis in Python}

\author{Leonardo Rydin Gorj\~ao}
\affiliation{Forschungszentrum J\"ulich, Institute for Energy and Climate Research - Systems Analysis and Technology Evaluation (IEK-STE), 52428 J\"ulich, Germany}
\affiliation{Institute for Theoretical Physics, University of Cologne, 50937 K\"oln, Germany}

\author{Galib Hassan}
\affiliation{Forschungszentrum J\"ulich, Institute for Energy and Climate Research - Systems Analysis and Technology Evaluation (IEK-STE), 52428 J\"ulich, Germany}
\affiliation{Institute for Theoretical Physics, University of Cologne, 50937 K\"oln, Germany}

\author{Jürgen Kurths}
\affiliation{Potsdam Institute for Climate Impact Research, 14473 Potsdam, Germany}
\affiliation{Institute of Physics, Humboldt University Berlin, 12489 Berlin, Germany}
\affiliation{Centre for Analysis of Complex Systems, World-Class Research Center ``Digital biodesign and personalised healthcare'', Sechenov First Moscow State Medical University, 119991 Moscow, Russia}

\author{Dirk Witthaut}
\affiliation{Forschungszentrum J\"ulich, Institute for Energy and Climate Research - Systems Analysis and Technology Evaluation (IEK-STE), 52428 J\"ulich, Germany}
\affiliation{Institute for Theoretical Physics, University of Cologne, 50937 K\"oln, Germany}

\begin{abstract}
Multifractal detrended fluctuation analysis (MFDFA) has become a central method to characterise the variability and uncertainty in empiric time series. 
Extracting the fluctuations on different temporal scales allows quantifying the strength and correlations in the underlying stochastic properties, their scaling behaviour, as well as the level of fractality. 
Several extensions to the fundamental method have been developed over the years, vastly enhancing the applicability of MFDFA, e.g. empirical mode decomposition for the study of long-range correlations and persistence.
In this article we introduce an efficient, easy-to-use \texttt{python} library for MFDFA, incorporating the most common extensions and harnessing the most of multi-threaded processing for very fast calculations.

\textbf{Software: \url{https://github.com/LRydin/MFDFA}}
\end{abstract}

\maketitle

\section{Introduction}
A common tool to unveil the nature of the scaling and fractionality of a process, natural or computer-generated, is Multifractal Detrended Fluctuation Analysis (MFDFA).
It was initially developed by Peng \textit{et al.}~\cite{Peng1994,Peng1995} as basic Detrended Fluctutation Analysis (DFA) and later extended to study multifractal processes by Kandelhardt \textit{et al.}, giving rise to MFDFA~\cite{Kantelhardt2002}.
It addresses the question of the presence of correlations in time series and can be employed to analyse both discrete as well as continuous-time stochastic processes.
Since its initial development in the late 90's, it has been revisited to incorporate several other elements, e.g. empirical mode decomposition as a method for detrending~\cite{Zhaohua2004,Zhaohua2009,Qian2011,Zhang2019}, overlapping moving windows~\cite{Zhou2010,Lai2019}, and a new metric denoted extended detrended fluctuation analysis~\cite{Pavlov2020a,Pavlov2020b,Pavlov2020c,Pavlov2021}.
There are several additional features exist, designed to study correlations of two or more time series~\cite{Podobnik2008, Zhou2008}, lag correlations in time series~\cite{Alvarez-Ramirez2009}, and Fourier-DFA~\cite{Chianca2005}, amongst others. 
A comprehensive study of DFA and the interplay between trends in data and correlated noise can be found in Ref.~\cite{Hu2001}.
MFDFA has found application in various fields, such as the analysis of heartbeat rate~\cite{Ivanov1999}, arterial pressure~\cite{Pavlov2020a}, EEG sleep data~\cite{Pavlov2020b,Pavlov2021}, physiology~\cite{Srimonti2013},  keystroke time series from Parkinson's disease patients~\cite{Madanchi2020}, cosmic microwave radiation~\cite{Movahed2011,Movahed2013}, seismic activity~\cite{Telesca2005,Shadkhoo2009}, sunspot activity~\cite{Movahed2006}, atmospheric scintillation~\cite{Tanna2014}, temperature variability~\cite{Meyer2019}, meteorology~\cite{Pedron2010}, precipitation levels~\cite{Tessier1996}, streamflow and sediment  movement~\cite{Matsoukas2000,Koutsoyiannis2003,Kantelhardt2006,Zhang2008,Rodriguez2013,Wu2018,Zhang2019}, protein folding~\cite{Figueiredo2010}, finance and econophysics~\cite{Zunino2008,Zunino2009,Grech2013,Drozdz2018,Minhyuk2018}, electricity prices~\cite{Weron2004b,Wang2013}, power-grid frequency~\cite{Shalalfeh2016,RydinGorjao2020}, epidemiology~\cite{Leung2011}, music~\cite{Jafari2007,Telesca2011,Ribeiro2012}, ethology~\cite{Alados2000,Rutherford2003}, multifractal harmonic signals~\cite{Li2019}, and microrheology~\cite{Madanchi2021}.

MFDFA is a numerical algorithm designed to determine the self-similarity of a stochastic process.
Putting it simply, the algorithm examines the relation between the diffusion of the process and its propagation in time or space.
Auto-regressive and stochastic processes with different power-law scaling will diffuse with different rates.
Fluctuation Analysis (FA) provides a method to uncover these correlations, but fails in the presence of trends in the data, which, for example, are particularly present in weather and climate data.
Detrending the data via polynomial fittings (DFA) allows one to uncover solely the relation between the inherent fluctuations and the time scaling of a process, thus circumventing the impact of non-stationarity in the data.
Likewise, other methods---as empirical mode decomposition or moving average windows---are viable options to detrend the data.
Another problem is that a process might be driven by more than one time scale, i.e., have more than one internal period, which can be removed either with local polynomial fittings or EMD.
Moreover, a stochastic process might be of a monofractal or multifractal nature.
By studying a continuum of power variations of DFA one extends into MFDFA, which permits the study of the fractality of the data by comparing power variations, i.e., a multifractal spectrum.

In this software we sought to design a computationally efficient code focused on computational speed and usability.
There are currently no flexible and available implementations of MFDFA in \texttt{python}. 
Available are some \texttt{MATLAB}~\cite{Ihlen2012} as well as \texttt{R} packages~\cite{Laib2018a,Laib2018b}.
There is a particularly thorough introductory guide to MFDFA in \texttt{MATLAB} with a source-code by Espen A. F. Ihlen~\cite{Ihlen2012}, which is easy to implement but numerically inefficient.
With this implementation efficiency was sought.
This was achieved by making the most out of \texttt{python}, reshaping the code to allow for multi-threading, especially relying on \texttt{numpy}'s \texttt{polynomial}, which scales easily with modern computers having more processor cores~\cite{NumPy}.
Moreover, this library contains the most commonly applied methods alongside with DFA and MFDFA: the added feature of empirical mode decomposition is implemented to substitute the polynomial fittings; A moving window is included, especially valuable for shorter time series; The extended DFA (eDFA) method is also included, adding a second metric of fractal scaling, especially valuable for multifractal or aperiodic time series.

In the following sections we will introduce MFDFA alongside some of the aforementioned methods incorported into the \texttt{MFDFA} library.
We will present two classical applications, one with a monofractal process and one with a multifractal noise, and show how to use MFDFA to extract their characteristics from a single one-dimensional time series.
\texttt{Python} code is presented to explicate the use of the \texttt{MFDFA} library.
Subsequently we study two real-world time series: the sunspot time series from 1818 to 2020 which accounts for the daily recorded sunspots and the quarter-hourly electricity trading market, which accounts for a small volume of electricity sell and purchase at 15 minute windows in Continental Europe.
Lastly we address a few details of the library and contribute a few closing remarks.

\section{Theoretical Basis}
In the following we briefly summarise the theoretical basis of Multifractal Detrended Fluctuation Analysis. 
Later we detail the different included extensions and which modifications these add to the original MFDFA algorithm.

\subsection{Multifractal Detrended Fluctuation Analysis}
Multifractal Detrended Fluctuation Analysis studies the variances of the fluctuations of a given process by considering increasing segments of a time series.

i) Take a time series $X(t)$ (in time or space $t$) with $N$ data points, discretised as $X_i$, $i=1,2, \dots, N$.
Find the ``detrended'' profile of the process by defining
\begin{equation}\label{eq:1}
  Y_i = \sum_{k=1}^i \left ( X_k - \mu_X \right),~\text{for}~i=1,2, \dots, N,
\end{equation}
i.e., the cumulative sum of $X_i$ subtracting the mean $\mu_X$ of the data.

ii) Section the data into smaller non-overlapping segments of length $s$, obtaining therefore $N_s = \text{int}(N/s)$ segments.
Given the total length of the data is not always a multiple of the segment's length $s$, discard the last points of the data.

iii) Consider the same data, apply the same procedure, but discard now instead the first points of the data.
One has now $2N_s$ segments of the time series.

iv) To each of this segments fit a polynomial $y_v$ of order $m$ and calculate the variance of the difference of the data to the polynomial fit
\begin{equation}\label{eq:F(v,s)}
  F(v,s) = \frac{1}{s} \sum_{i=1}^s [Y_{(v-1)s + i} - y_{(v-1)s + i}]^2,
\end{equation}
for $v=1,2, \dots, N_s$, where $y_{(v-1)s + i}$ is the polynomial fitting for the segment $Y_{(v-1)s + i}$ of length $s$, fitted via least-squares.
The order of the polynomial $y_v$ can be freely chosen, giving rise to the denotes (MF)DFA1, (MF)DFA2, $\dots$, (MF)DFA$m$, dependent on the chosen degree $m$ of the polynomial.

v) Notice now $F(v,s)$ is a function of each variance of each $v$-segment of data and of the different $s$-length segments chosen.
Define the $q$-th order fluctuation function by averaging over the $N_s$ variances of the segments of size $s$
\begin{equation}\label{eq:F}
  F_q(s) = \left\{\frac{1}{N_s} \sum_{v=1}^{N_s} [F(v,s)]^{q/2}\right\}^{1/q}.
\end{equation}
The fluctuation function $F_q(s)$ depends on two parameters: the segment size $s$ and the $q$-th power.
The fluctuation function $F_q(s)$ is the function we will focus on which the MFDFA algorithm developed extracts from the data.

Two closely related algorithms are discussed and introduced here, DFA~\cite{Peng1994} and MFDFA~\cite{Kantelhardt2002}.
DFA is a particular case of MFDFA for the choice of $q=2$.
What is presented above is the MFDFA algorithm as according to Kantelhardt \textit{et al.}~\cite{Kantelhardt2002}, for which a particular choice of $q=2$ leads to the fluctuation function $F_2(s)$.
The DFA fluctuation function $F_2(s)$ can unveil solely the monofractal spectrum of a time series.
If the examined time series $X_i$ is monofractal, DFA is sufficient to describe and uncover the scaling relations in the data.
If not, one must rely on MFDFA and the study of the spectrum unveiled by varying the $q$-th power.

We will later detail two changes: i) The first involving empirical mode decomposition (EMD) for detrending, where the local polynomial fittings are replaced and the trends of the data are subtracted by removing select Intrinsic Mode Functions (IMFs) obtained via empirical mode decomposition.
ii) The second change involves substituting the non-overlapping segments with overlapping ones.

\begin{figure}[t]
\includegraphics[width=\linewidth]{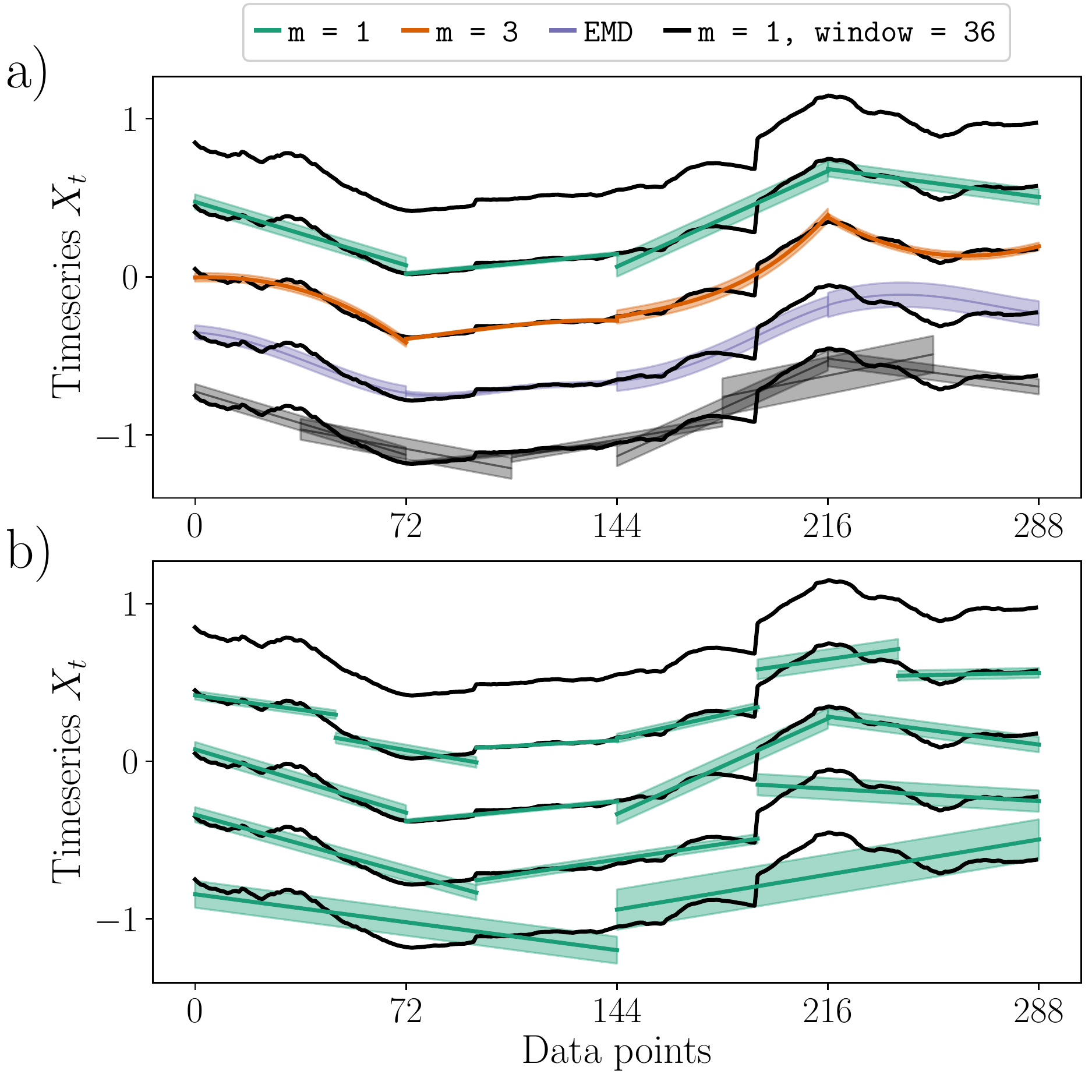}
\caption{Multifractal Detrended Fluctuation Analysis (MFDFA) of an exemplary $288$ data points time series $X_t$.
Panel a) shows, from top to bottom: the time series; first-order polynomial fit (\texttt{m = 1}); third-order polynomial fit (\texttt{m = 3}); EMD detrending with the slowest Intrinsic Mode Function (\texttt{EMD}); first-order polynomial fit (\texttt{m = 1}) with a moving windows with a step of $36$ data points (\texttt{window = 36}).
Segments with a size of $s=72$ data points.
The lines indicate the fits, either via polynomials or EMD.
Panel b) displays the changing segment size $s$ for a first-order polynomial fit (\texttt{m = 1});
From top to bottom: the time series; segmentation with $s=48$; $s=72$; $s=96$; $s=144$. 
Shaded areas on both panels indicated the standard deviation of each segment.}\label{fig:sketch}
\end{figure}

The inherent scaling properties of the data, if the data displays power-law correlations, can now be studied in a log-log plot of $F_q(s)$ versus $s$, where the scaling of the data obeys a power-law with exponent $h(q)$ as
\begin{equation}\label{eq:scaling}
  F_q(s) \sim s^{h(q)}
\end{equation}
where $h(q)$ is the generalised Hurst exponent or \textit{self-similarity} exponent, which will dependent on $q$ if the data is multifractal, and relates directly to the Hurst index~\cite{Hurst1951}.
The generalised Hurst exponent $h(q)$ is obtained by finding the slope of $F_q(s)$ curve in the log-log plots.

If the data is monofractal, the generalised Hurst exponent $h(q)=H$ is independent of $q$ and the generalised Hurst exponent is simply the Hurst index $H$.
On the other hand, if the data is multifractal, the dependence on $q$ can be understood by studying the multifractal scaling exponent $\tau(q)$, given by
\begin{equation}
    \tau(q) = qh(q) - 1,
\end{equation}
which depends on the generalised Hurst exponent $h(q)$.
Similarly, one can construct the singularity spectrum $D(\alpha)$ as the Legendre transform~\cite{Hentschel1983,Halsey1986,Kurths1987,Meneveau1987}. If $\tau(q)$ is sufficiently smooth, the singularity strength $\alpha$ is given by
\begin{equation}
    \alpha = \tau'(q) = h(q) + qh'(q),
\end{equation}
from which the singularity spectrum $D(\alpha)$ can be constructed as
\begin{equation}\label{eq:sing_spect}
    D(\alpha) = q \alpha - \tau(q).
\end{equation}

The singularity spectrum $D(\alpha)$ describes the dimension of the subset of the time series which is characterised by the singularity strength $\alpha$~\cite{Salat2017}.
The breadth of singularity strength $\alpha$ indicates the strength of the multifractality of the time series, centred around the most prominent scale of the time series, i.e., $h$.
The singularity spectrum $D(\alpha)$ takes the shape of an inverted parabola with a maximum at $D(\alpha=0)=D_0$, known as the box-counting or Minkowski--Bouligand dimension, or sometimes simply fractal dimension~\cite{Hentschel1983}.
$D(\alpha=1)=D_1$ is known as the information dimension and $D(\alpha=2)=D_2$ the correlation dimension~\cite{Falconer2004}.
For a clearer discussion of these properties, see Refs.~\cite{Barabasi1991,Kantelhardt2002}. 
An extensive and very illustrative representation of this can be found in Ref.~\cite{Ihlen2012}.
For a careful analysis of the meaning and interpretation of the generalised Hurst coefficients extracted from (MF)DFA, see Ref.~\cite{Serinaldi2010}, where a description and clarification is given on what are persistent and anti-persistent motions, stationary and non-stationarity time series, among other relevant details.

\subsubsection{Empirical mode decomposition}
Empirical mode decomposition (EMD) is a method with a variety of applications in time series analysis~\cite{Huang1998}.
It seeks to extract the modes of oscillation of a time series strictly from the data.
One can harness the ability of the EMD, i.e., the Hilbert--Huang decomposition of a time series, to obtain the trend or trends of the time series and utilise those to transform non-stationary into stationary data.
The central concept, developed by Qian, Gu, and Zhou~\cite{Qian2011}, is to substitute the detrending method employed in the traditional MFDFA, i.e., polynomial fittings, by removing instead particular Intrinsic Mode functions extracted via EMD.
A sketch of the method can be seen in Fig.~\ref{fig:sketch}.

EMD can be summarised in a few steps: a set of intrinsic mode functions (IMFs) are extracted from the time series, obeying: 1) the number of extrema and the number of zero crossings must maximally differ by one. 2) for any point, the mean value of the envelope defined by the local extrema is zero.
Numerical methods---as cubic splines---are used to find the curve that best fits ``between'' the local extrema of the time series.
The method is applied iteratively: i) Obtain an IMF by finding the ``best'' curve between the local extrema of the time series; ii) Subtract this IMF to the time series; iii) Repeat.
Apply the process recursively to the time series until the final IMF contains solely a residual trend of the data.

\subsubsection{Moving windows}
The overlapping moving windows included in this library is not aimed at detrending, but instead for the analyses of rather short time series or very large scales in longer time series~\cite{Zhou2010}.
In the literature several applications of moving average windows have been proposed as methods to remove trends and ensure stationarity, by simply removing a windowed average to the time series~\cite{Zhou2010,Minhyuk2018,Lai2019}.
This is not what we do here.
Here, we substitute the non-overlapping segmentation, as explain after Eq.~\eqref{eq:1}, by a moving window, replacing the two separate segmentations by a moving window of each segment size $s$ over the time series, as proposed by Zhou and Leung~\cite{Zhou2010}.
This is particularly relevant when examining short time series, where quickly the choice of larger lags $s$ separates the data into a small number of segments, resulting in a poor statistics for the scaling at larger lags.
A sketch of the methods can be seen in Fig.~\ref{fig:sketch}.

\subsubsection{Extended Detrended Fluctuation Analysis}

A new metric of similar nature as the fluctuation function $F_q(s)$, given in Eq.~\eqref{eq:F}, has been proposed in Ref.~\cite{Pavlov2020a}
This measure supersedes the $q$-order powers and takes in solely the case of DFA where $q=2$.
Instead of finding the average of the variances over each choice of segments of size $s$, it considers the difference between the extrema of the fluctuation function at each segment $s$.
Take $F(v,s)$ as given in Eq.~\eqref{eq:F(v,s)} and extract the maximum and minimum of the variances over all windows $v$ for a certain window size $s$
\begin{equation}
       \Delta F(s) = \max_v[F(v,s)] - \min_v[F(v,s)].
\end{equation}
This new metric $\Delta F(s)$ is denoted Extended Fluctuation Analysis.
In general, $\Delta F(s)$ can scale as a power law with a different exponent
\begin{equation}
       \Delta F (s) \sim s^\beta.
\end{equation}
This metric takes into account aperiodicities in the data which, in some sense, would be accounted for as a multifractal behaviour.
It can unravel a second scaling phenomenon due to local changes of a time series' period.

\section{Examples}

To exemplify the usage of MFDFA, we first take two common examples of stochastic processes, a fractional Ornstein--Uhlenbeck process and general process that has a symmetric Lévy $\alpha$-stable distribution, with single parameter $\alpha$.
We will show how to extract the fluctuation function $F_q(s)$ and how to interpret the plots conventionally extracted to perform the analysis.
Subsequently we test the algorithm with real-world data on sunspot time series, following Ref.~\cite{Movahed2006}, and later apply the algorithm to electricity price time series from the European Power Exchange.

\subsection{Numerically generated data}

\subsubsection{Fractional Ornstein--Uhlenbeck process}

\begin{figure*}[t]
\includegraphics[width=\linewidth]{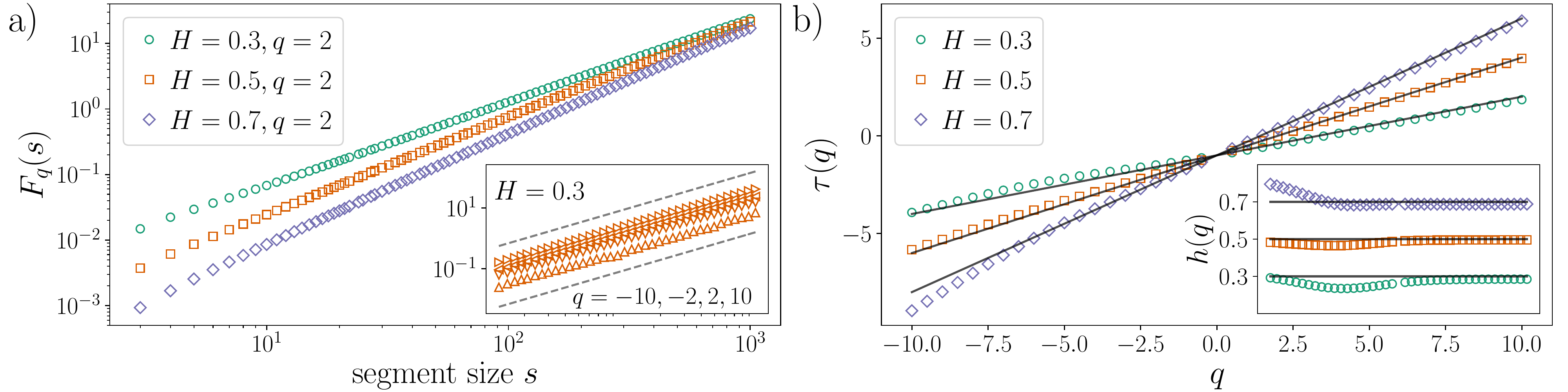}
\caption{Multifractal Detrended Fluctuation Analysis (MFDFA) of three exemplary sample paths of fractional Ornstein--Uhlenbeck processes, given by Eq~\eqref{eq:OU}, with Hurst indices of $H=0.3, 0.5$, and $0.7$.
Panel a) displays the log-log plot of the segment size $s$ versus the fluctuation function $F_2(s)$, given by Eq.~\eqref{eq:F}, for $q=2$.
Each line has a slope of $H+1$, as expected.
The inset shows $F_q(s)$ for the case of $H=0.3$ and the power variations $q=-10,-2,2,10$.
These lines are all parallel indicating that the process is monofractal, as expected.
The dashed lines indicate the theoretical expected scaling, i.e., a slope of $H+1 = 0.3 + 1$, where the $+1$ account for the increase in regularity due to the integration.
The generalised Hurst coefficients $h(q)$, which are simply $H+1$, are obtained by extracting the slopes of the curves (in a log-log scale).
Panel b) shows the multifractal scaling exponent $\tau(q)$, given by Eq.~\eqref{eq:scaling}, which exhibits a linear dependency, i.e., $h(q)=H$, indicating again the process is monofractal.
The processes were numerically integrated with an integration step $\Delta t = 0.001$ over $N=10^4$ time units ($N=10^7$ data points).
The \texttt{MFDFA} algorithm ran in $1$\,min $29$\,s $\pm$ $1.85$\,s, for $100$ segments $s$ and $40$ $q$-variation powers, with first-order polynomial fits.}\label{fig:OU}
\end{figure*}

To study the scaling effects in continuous stochastic processes, three exemplary fractional Ornstein--Uhlenbeck processes are taken, defined as~\cite{Tabar2019}
\begin{equation}\label{eq:OU}
  dX_t = - \theta X_t dt + \sigma d B^H_t,
\end{equation}
with a fractional Brownian motion $B^H_t$ with the covariance function
\begin{equation}\label{eq:BM}
  \mathbb{E}\left[B^H_tB^H_{t'}\right] = \frac{1}{2}\left(|t|^{2H} + |t'|^{2H} - |t-t'|^{2H}\right).
\end{equation}
Eq.~\eqref{eq:OU} fixed mean reverting strength $\theta=1.0$ and volatility $\sigma=0.5$, with three distinct Hurst indices of $H=0.3$, $0.5$, and $0.7$. 
Note here that the classic uncorrelated Brownian motion is given by $H=0.5$.
A fractional Brownian motion has a self-similarity exponent given by the Hurst index $H$, thus the three choices of fractional Ornstein--Uhlenbeck should result in a scaling of $h(q) = H + 1$.
The $+1$ is due to the integration, which smooths the fluctuations and thus increases the regularity of the process.
We will now numerically integrate these processes and utilise the \texttt{MFDFA} library to identify the Hurst coefficients and the presence of a monofractal vs multifractal spectrum in the time series.

Let us exemplify how to numerically generate data and utilise the \texttt{MFDFA} library
Load the \texttt{MFDFA} library alongside with the fractional Brownian noise generator \texttt{fgn} included in your \texttt{python} console or editor.

\begin{lstlisting}[language=Python, caption=Load the \texttt{MFDFA} library]
from MFDFA import MFDFA
from MFDFA import fgn
\end{lstlisting}

\noindent To numerically integrate an Ornstein--Uhlenbeck process, given by Eq.~\eqref{eq:OU}, we utilise an Euler--Maruyama scheme with a stepsize $\Delta t = 0.001$ for a total time of $t=10^4$ (thus we have $10^7$ data points).
Here exemplified is the fractional Ornstein--Uhlenbeck process with $H=0.3$.

\begin{lstlisting}[language=Python, caption=Integrate Ornstein--Uhlenbeck process, firstnumber=3]
# integration time and time sampling
t_final = 10000
delta_t = 0.001
N = int(t_final/delta_t)

# The parameters theta and sigma
theta = 1
sigma = 0.5

# Initialise the array X
X = np.zeros(N)

# Generate the fractional Brownian noise
# with a Hurst coefficient of H = 0.3
dB = (t_final ** H) * fgn(N, H = 0.3)

# Integrate the process
for i in range(1,N):
    X[i] = X[i-1] - theta*X[i-1]*delta_t + sigma*dB[i]
\end{lstlisting}

\noindent To retrieve the MFDFA spectrum of the generated time series, define the set of $q$ power variations and the lags $s$ to examine, and call the \texttt{MFDFA} function.

\begin{lstlisting}[language=Python, caption=Applying \texttt{MFDFA}, firstnumber=22]
# 100 lag s points from 3 to 1000
lag = np.logspace(0.6,3,118).astype(int)
lag = np.unique(lag)

# q power variations, removing 0 power
q = np.linspace(-10,10,41)
q = q[q!=0.0]

lag, fluct = MFDFA(X, lag = lag, q = q)
\end{lstlisting}

\noindent When not declaring the values of the $q$ powers, $q=2$ is assumed, thus resulting in the conventional DFA.
Likewise, not declaring the order of the polynomial fitting, a first-order polynomial is assumed, i.e., \texttt{order = 1}.

In Fig.~\ref{fig:OU} the MFDFA of the three processes can be seen.
In panel a) the fluctuation function $F_2(s)$, with $q=2$, is shown for a polynomial detrending of first order.
This is the conventional DFA.
The slopes of each curve in the log-log plot reveal the Hurst indices of each process, i.e., the fractional Ornstein--Uhlenbeck with Hurst $H=0.3$ scales with a slope of $1.3 = 0.3 + 1$, the other two with $H=0.5$ and $H=0.7$ have a slope of $1.5$ and $1.7$, respectively.
The inset in panel a) shows the fluctuation function $F_q(s)$, with $q=-10$, $-2$, $2$, and $10$, for the fractional Ornstein--Uhlenbeck process with $H=0.3$.
Note that the slope of all power variations is the same, i.e., the process is monofractal, as expected.
The monofractality of the process is also evident in panel b).
The multifractal scaling exponent $\tau(q)$ is shown and is purely a linear function.
Likewise, in the inset, the generalised Hurst indices $h(q)$ for the three processes for a set of power variations $q\in[-10,10]$ is displayed.
The linear shape of $\tau(q)$ and constant value of $h(q)$ in the inset indicates, as expected, that these three processes are monofractal.
Small deviations are seen for very negative $q$ powers ($q\lesssim7$), which arise due to the numeric (negative) powering operation, which highly depends on the numerical precision of the data.

\begin{figure*}[t]
\includegraphics[width=\linewidth]{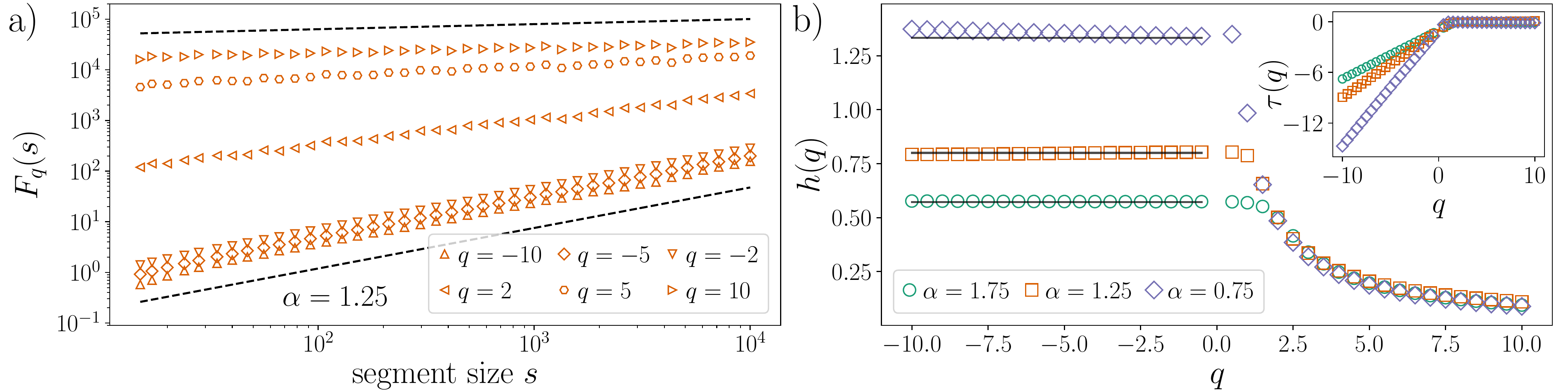}
\caption{Multifractal Detrended Fluctuation Analysis (MFDFA) of three exemplary symmetric Lévy $\alpha$-stable distributed processes, with $\alpha = 1.75$, $1.25$, and $0.75$.
In panel a) the fluctuation function $F_q(s)$ is shown as a function of the segment size $s$ on  double logarithmic scales for $\alpha = 1.25$ and different values of the power, $q = -10, -5, -2, 2, 5, 10$.
For $q>\alpha$ the curves are not parallel, indicating the multifractal nature of the process.
Panel b) displays the generalised Hurst exponent $h(q)$, where a clear non-linear dependency on $q$ is observable.
The inset displays the multifractal scaling exponent $\tau(q)$ displaying two clear distinct behaviours for $q<0$ and $q>\alpha$.
The solid lines indicate the theoretical expected scaling for $q<0$.
The three processes were drawn from Lévy $\alpha$-stable distributions, each with $N=10^7$ data points.
The \texttt{MFDFA} algorithm ran in $1$\,min $24$\,s $\pm$ $2.17$\,s for $100$ segments $s$, $40$ $q$ powers, and third-order polynomial fits.}\label{fig:Lévy}
\end{figure*}

\subsubsection{Lévy-driven process}

As a second example, take a collection of L\'evy distributed random variables~\cite{Applebaum2011}.
That is, each $X_t$ is drawn independently from a symmetric $\alpha$-stable distribution, such that the probability density function of $X(t)$ vanishes as a power-law $P(x)\sim |x|^{-(\alpha-1)}$ for large $|x|$~\cite{Applebaum2011}. 
These processes exhibit heavy tails, ill-defined variances, and multifractal scaling. 
In Fig.~\ref{fig:Lévy} three symmetric Lévy $\alpha$-stable distributed processes with $\alpha = 1.75, 1.25$, and $0.75$ are studied with MFDFA.

The multifractal behaviour can be identified directly in panel a), where the fluctuation function $F_q(s)$ for $\alpha = 1.25$ is shown. 
The lines of $F_q(s)$ are not parallel for different positive $q$ powers, showing that the process is not mono-fractal. 
In fact, the process is bi-fractal, having a separate behaviour for $q<0$ and $q>\alpha$.
For positive power variations $q>\alpha$, the generalised Hurst exponent $h(q)$ decays like $1/q$.
For values of $q<0$, the generalised Hurst exponent $h(q) = 1/\alpha$.
This can be seen clearly in Fig.~\ref{fig:Lévy} b), where the generalised Hurst exponent $h(q)$ is displayed.
In the inset, one notices that the multifractal scaling exponent $\tau(q)=0$, for $q>\alpha$ (always zero for $q>2$), once again showing that none of these processes are distinguishable for positive power variations.

In general, without the aid of the multifractal spectra, which we uncovered by studying MFDFA for a range of $q$ values, it is not possible to distinguish between Lévy distributed processes.
The particular choice of $q=2$, i.e., conventional DFA, obscures the fractality of these processes, as they all show a similar scaling for $q=2$, i.e., $h(q=2)=1/2$ for all Lévy motions, including (non-fractional) Brownian motions (where $\alpha=2$).

\subsection{Real-world data, empirical mode decomposition, and extended DFA}\label{sec:real_world_data}

In order to evaluate the efficiency of the algorithm, we test here two real-world data sets.
Firstly, using MFDFA we will evaluate the multifractality of sunspots time series, a recurring phenomenon on the Sun's photosphere which can be observed with a telescope~\cite{Movahed2006}.
Secondly, we will analyse the German and Autrian spot market intraday quarter-hourly electricity price extracted from the European Power Electricity Exchange (EPEX SPOT)~\cite{EPEX,PriceData}.
We illustrate the application of two advanced features of the developed \texttt{python} package, moving windows and the masking of missing data points.

\subsubsection{Sunspots}
\begin{figure*}
\includegraphics[width=\linewidth]{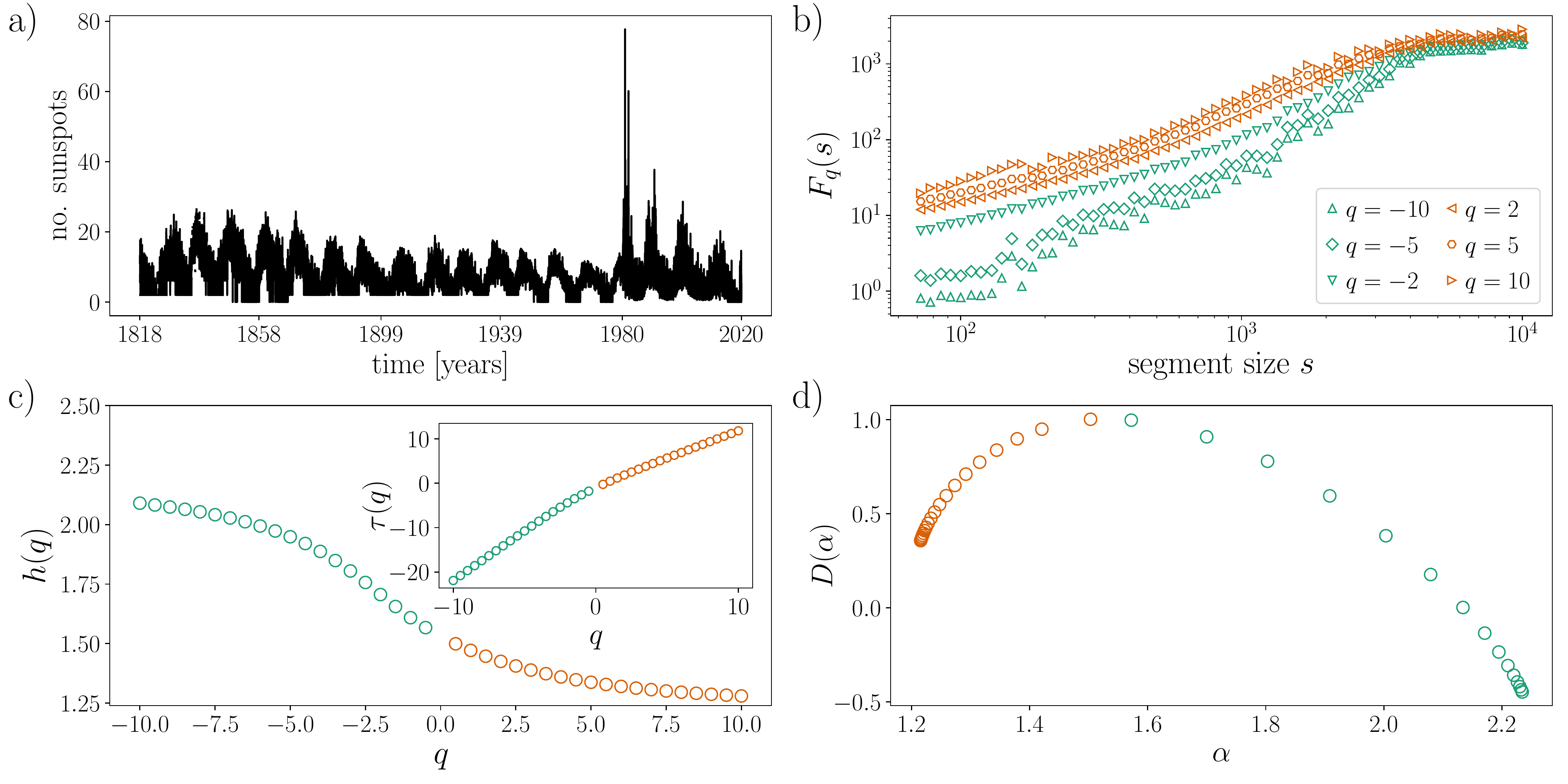}
\caption{Multifractal Detrended Fluctuation Analysis (MFDFA) of sunspot time series from 1818 to 2020, by the ILSO World Data Center, Royal Observatory of Belgium, Brussels~\cite{SunspotData}.
Panel a) shows the number of sunspots registered from 1818 to 2020.
Panel b) displays the fluctuation function $F_q(s)$ as a function of the segment size $s$ on a double-logarithmic scale for $q=-10,-5,-2,2,5,10$, with positive $q$ values in orange and negative $q$ values in green.
Panel c) displays the generalised Hurst coefficient $h(q)$ over $q$, and the inset displays the multifractal scaling exponent $\tau(q)$, given by Eq.~\eqref{eq:scaling}, both  highlighting the multifractal spectrum of the data ($h(q)$ is not constant over $q$, $\tau(q)$ is not linear over $q$).
Panel d) displays the singularity spectrum $D(\alpha)$ over the singularity strength $\alpha$ which shows a large breadth of $\alpha$ spanning over $[1.25,2.25]$, indicating the strong multifractality of the data.
The \texttt{MFDFA} algorithm ran in $426$\,ms $\pm$ $11.8$\,ms, for $70$ segments $s$, $40$ $q$ powers, and first-order polynomial fits.
The missing values were neatly removed by utilising \texttt{numpy}'s \texttt{masked} arrays, which is integrated in \texttt{MFDFA} and allows the user to simply ``mask'' empty or corrupted data.}\label{fig:Sunspots}
\end{figure*}

The sunspot numbers, also called Wolf numbers, are a rather simple measure of solar activity by counting in a weighted manner the number of groups of sunspots and single sunspots visible from the Earth in the solar photosphere, i.e. it is an integrated measure over space~\cite{Stix2002,Andre2015}.
Hence, the sunspot numbers form a time series which has a mean period of about 11 years, but is far from being simply periodic~\cite{Kurths1987}.
Solar activity is the result of complex magneto-hydrodynamic processes in the Sun characterised by a highly complex spatio-temporal dynamics.
It is of special interest to analyse the rather long series of sunspot numbers in order to explore some relations to the underlying spatio-temporal system.

The emergence of sunspots has a distinct statistics and a multifractal spectrum which has been examined in Ref.~\cite{Movahed2006}.
This publication has become a reference for multifractal studies as the data from the ILSO World Data Center, Royal Observatory of Belgium, Brussels is freely available~\cite{SunspotData}.
Here, we will focus on numerical efficiency and how to deal with missing or corrupt data.
We utilised another feature integrated in \texttt{MFDFA} that enables an efficient management of missing data points.
In \texttt{python}'s \texttt{numpy} arrays, missing or corrupt values in a time series can be handled with \texttt{masked} data, which logs the missing data points and takes these into account while performing averages, sums, and power operations.
When calculating averages or the variance of a segment, or when taking powers, the masked entries are not taken into account.
For the particular application with sunspot time series, which are recorded daily since 1818, there are $3247$ missing values, over a total of $74145$ entries, i.e., roughly $4.4\%$ of the data is missing.
To go around this, simply use

~%

\begin{lstlisting}[language=Python, caption=\texttt{MFDFA} and missing data, firstnumber=1]
# Read data whichever way preferred
data = read_data('sunspot.csv')

# Mask missing values. For this case -1
# is a missing entry in the record
data[data ==-1.] = np.nan
data = np.ma.masked_invalid(data)

# Run MFDFA (choose lag and q)
lag, fluct = MFDFA(data, lag, q)
\end{lstlisting}

\noindent The \texttt{MFDFA} will extract the variances as it is possible, taking into account the missing values in the time series.
Here we highlight that \texttt{MFDFA} calculated $40$ $q$-powers over $70$ segments $s$ in $426$\,ms $\pm$ $11.8$\,ms.

In Fig.~\ref{fig:Sunspots} we display the fluctuation function $F_q(s)$ for $q=-10,-5,-2,2,5,10$, in panel a), for $s\in[70,1000]$
These curves are not parallel, suggesting the time series is not monofractal.
In panel b) the generalised Hurst exponent $h(q)$ is shown as function of $q$, and similarly the multifractal scaling exponent $\tau(q)$ in inset.
The generalised Hurst exponent $h(q)$ is not constant over $q$ and consequently the multifractal scaling exponent $\tau(q)$ is not linear, indicating clearly the time series in multifractal.
In panel c) we display the singularity spectrum $D(\alpha)$ over the singularity strength $\alpha$, as given by Eq.~\eqref{eq:sing_spect}.
The singularity strength $\alpha$ spans a wide range of values, over $[1.25,2.25]$, indicating how strongly multifractal the time series is.
Here recall that a monofractal time series, as the fractional Ornstein--Uhlenbeck previously shown in Fig.~\ref{fig:OU}, has a very narrow range of the singularity strength $\alpha$, centred around $H$.
For the case of sunspot time series we see a wide range of $\alpha$, indicating the various scales of the phenomenon.
Note as well that $h(q)$ and $\alpha$ are always larger than $1$, indicating that this is a non-stationary process.

\subsubsection{German and Austrian spot market intraday quarter-hourly electricity price time series}
\begin{figure*}
\includegraphics[width=\linewidth]{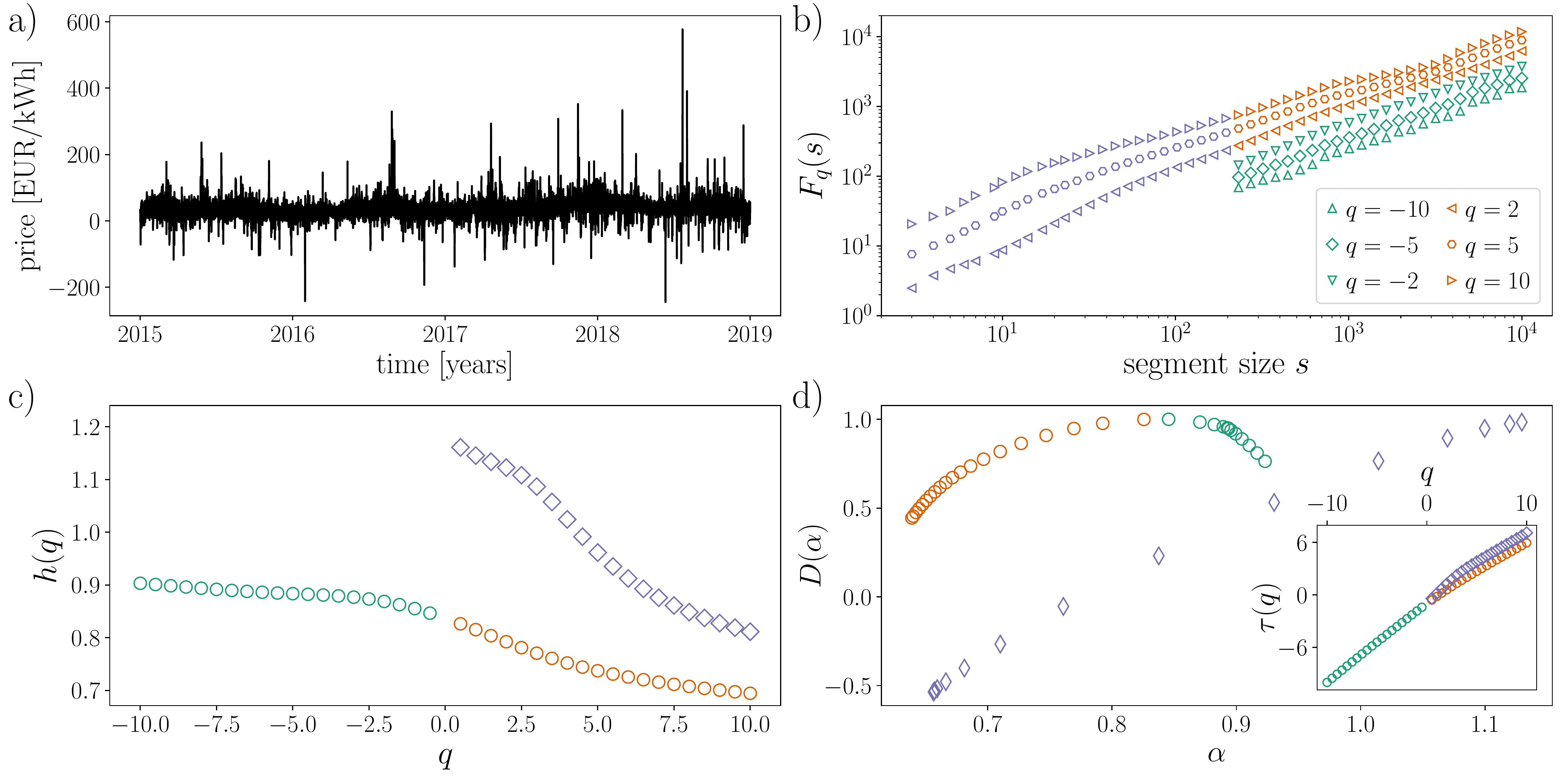}
\caption{Multifractal Detrended Fluctuation Analysis (MFDFA) of the spot market intraday quarter-hourly German and Austrian electricity price time series from 2015 to 2019, traded at the European Power Exchange (EPEX SPOT)~\cite{EPEX,PriceData}.
Panel a) displays the price in EUR/kWh from 2015 to 2019.
Panel b) displays the log-log plot of the segment size $s$ versus the fluctuation function $F_q(s)$ for $q=-10,-5,-2,2,5,10$.
Orange and green markers indicate the segments larger than two days, where purple indicate segments between 1 and 48 hours.
There two scales are studied separately.
Panel c) displays the generalised Hurst coefficient $h(q)$ over $q$, and the inset displays the multifractal scaling exponent $\tau(q)$, given by Eq.~\eqref{eq:scaling}.
Panel d) displays the singularity spectrum $D(\alpha)$ over the singularity strength $\alpha$.
The short-time scale (1--48 hours) displays large generalised Hurst coefficient $h(q)$ and a very large breadth of the singularity strength $\alpha$, indicating precisely the high volatility of the market at short time scales.
In comparison, the longer time scales ($>48$ hours) are much ``milder'', and the variations of $\alpha\in[0.67,0.92]$, which indicates the process is both stationary, int the long run, and only moderately volatile.
The \texttt{MFDFA} algorithm ran in $2$\,min $11$\,s $\pm$ $4.43$\,s for $50$ segments $s$, $40$ $q$ powers, first-order polynomial fits, and the moving window.}\label{fig:eex}
\end{figure*}

We will examine now a 4-year long time series sampled at 15 minutes of the spot market intraday quarter-hourly German and Austrian electricity price~\cite{EPEX,PriceData}, from the 1st of January 2015 to end of December 2019, traded at the European Power Exchange (EPEX SPOT).
To the extent of our knowledge no multifractal analysis of this particular data has been performed before, but other multifractal studies of price time series exist~\cite{Wang2013}.
In Wang \textit{et al.}~\cite{Wang2013}, the authors examine different scaling properties for selected periods of low, regular, and high electricity price for some United States of America's electricity markets in the year 2000 and 2001.
Here we propose a different analysis, studying the data and examining a short and long time scale of the data without separating different activity periods.

We know that the 15 minute trading electricity market amounts to a small volume of the overall exchange electricity sold, thus this market serves only electricity producers which can either extract or inject power from the power-grid system in a very fast manner ($<15$ minutes)~\cite{Braun2018,Narajewski2020}.
This will lead us to explore to separate scaling phenomena in the data: A short and a long timescale of market activities.
The expectation is that the very short-time trading is highly volatile, given the necessity of the power grid in injecting or extracting power is a fairly speedy manner.
In the long run, the quarter-hourly market is intrinsically linked to the larger hourly and daily electricity market, which has far less variability, as most of the power is sold in lengthier contracts, stabilising the value of the electricity price.
Thus one expects a narrower multifractality at large temporal scales.
Multifractality is nevertheless expected, as the system exhibits very large yet seldom negative prices, as well as an occasional four of five-fold increase of the (positive) prices, again occurring seldom and lasting very short periods.

In order to obtain a better statistics of the shorter time scales, we will employ \texttt{MFDFA}'s moving windows method previously discussed.
The moving window method requires the input of the number of steps used to ``move'' the windows.
For the following example, the \texttt{window} parameter is set to \texttt{1}, thus each overlapping window is displaced by solely $1$ data point.
This substantially increases the computational time as each averaging operation is repeated by the number of segments $s-1$.
For Fig.~\ref{fig:eex}, the total calculation lasted $2$\,min $11$\,s $\pm$ $4.43$\,s for the windowed mode, compared with $764$\,ms $\pm$ $9.32$\,ms with the conventional non-overlapping windows, for $50$ $s$ segments.

\begin{lstlisting}[language=Python, caption=\texttt{MFDFA} and moving window, firstnumber=1]
# Read data whichever way preferred
data = read_data('price.csv')
	
# Run MFDFA (choose lag and q)
lag, fluct = MFDFA(data, lag, q, 
	extension = {'window' : 1})
\end{lstlisting}

In Fig.~\ref{fig:eex} we display the MFDFA analysis of the price time series, as previously done for the sunpots in Fig.~\ref{fig:Sunspots}.
We perform a similar analysis as above, thus we will condense the technical details and focus on the interpretation.
Previous studies point to a clear separation of the scaling behaviour of price time series~\cite{Simonsen2003,Weron2004,Weron2004b}. 
They separate two time scales for periods shorter and longer than 24 hours.
These studies used pricing data from before 2004 for the Nordic grid (Nordpool).
In our analysis, we similarly separate two scales, between 1 and 48 hours and between 48 hours and 10 days.
These are indicated in purple (short timescale) and orange and green (long time scale).
We first observe that negative $q$ powers do not exist for the short time scale.
This is not unusual, many processes do not show a multifractal spectrum with negative $q$ values.
Note that this involves taking negative powers of the average of the variances, which is not always well defined for short $s$ segments.
This also served as a threshold to assess the change in the fractal behaviour of the time series.
For the large time scales ($>48$ hours) the negative powers are well defined, and we can identify the full singularity spectrum $D(\alpha)$, as seen in panel c).
We note that the short time scale ($<48$ hours) has a very strong multifractality (in purple).
The singularity strength $\alpha$, which we can only extract for positive $q$ values, has a very large breadth, especially with its equivalent for the large time scale (in orange).
This is well grounded on the previous arguments of having a very volatile market at these short time scales, thus these results are in line with what is known about this market: The high volatility and occasional burst---into very large electricity prices or into negative prices---generate a wide range of the singularity strength.
The long-term stability, connected with the larger intraday and day-ahead electricity markets, makes the process far less multifractal at large  temporal scales.

\section{The \texttt{MFDFA} library}
The Multifractal Detrended Fluctuation Analysis library \texttt{MFDFA} in \texttt{python} presented is a standalone package based integrally on \texttt{python}'s \texttt{numpy}~\cite{NumPy}.
It can be found in \url{https://github.com/LRydin/MFDFA}.
It harnesses \texttt{numpy}'s vectorised polynomial fittings, making it possible to utilise all computational cores in a computer's processor(s). 
Additionally, EMD is included as an extra feature, which is integrated into \texttt{MFDFA} by simply installing the \texttt{python} library \texttt{PyEMD}~\cite{Laszuk2017}. 
The conventional plots associated with multifractal analysis, i.e., $F_q(s)$ \textit{vs.} $s$, $h(q)$ and $\tau(q)$ \textit{vs.} $q$, and $D(\alpha)$ \textit{vs.} $\alpha$, are available as well and require the plotting library \texttt{matplotlib}~\cite{Matplotlib}.

The \texttt{MFDFA} library accepts \texttt{numpy}'s \texttt{masked} arrays, which is particularly convenient when dealing with time series with missing data, as exemplified in Sec.~\ref{sec:real_world_data} and Fig.~\ref{fig:Sunspots}.

\begin{figure}
\includegraphics[width=1\linewidth]{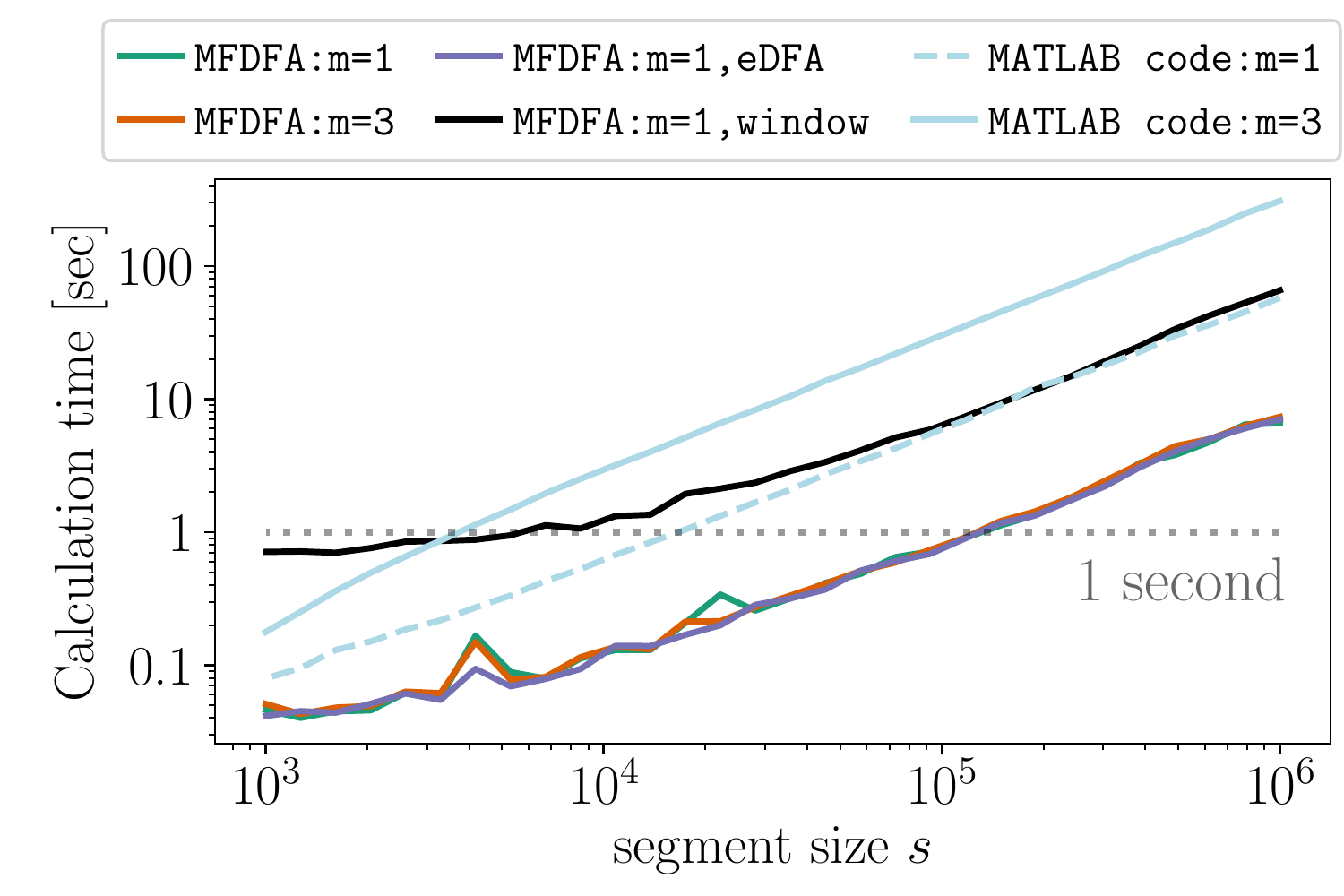}
\caption{Speed performance of \texttt{MFDFA} for time series with sizes varying between $[10^3, 10^6]$ data points of a fractional Ornstein--Uhlenbeck as given by Eq.~\eqref{eq:OU}.
Included are first-order and third-order polynomial fits, first-order fits with extended DFA, and first-order fits with a moving window with a step size of $5$.
A comparison with the distributed \texttt{MATLAB} code is included~\cite{Ihlen2012}.
Tests ran on \texttt{python} 3.8.2 and \texttt{MATLAB}~R2020b.
\texttt{MFDFA} has a average speed-up compared with the \texttt{MATLAB} code, with a five-fold speed increase for first-order polynomial fits (\texttt{m=1}) and a $\times27$-fold increase for third-order polynomials fits (\texttt{m=3}).
Both codes were tested for $100$ segments $s$ and $40$ $q$ powers.
All tasks were performed on a laptop on two computer cores at 2.9\,GHz each.}\label{fig:performance}
\end{figure}

The \texttt{MFDFA} library offers a considerable speed-up in comparison with the available \texttt{MATLAB} version.
The library is fully developed to work with multi-threading, which shows an increase in the performance, while handling time series larger than $10^5$ data points.
In Fig.~\ref{fig:performance} we display the performance of the \texttt{MFDFA} library for time series of fractional Ornstein--Uhlenbeck processes given in Eq.~\eqref{eq:OU} of increasing length.
The MFDFA operation scales linearly with the number of points of the generated time series.
The MFDFA algorithm runs in under $1$ second for time series having up to $10^5$ datapoints, with a first-order polynomial fittings, $100$ segments $s$ and $40$ $q$ powers, and outperforms the conventional library in \texttt{MATLAB} by up to a factor of $\times10^3$ in computational speed.

Estimation error and significance calculation have not been included in the library, as the focus lied on computational speed and the inclusion of several extra features, as discussed.

\section{Conclusion}
We have presented a numerically efficient \texttt{python} implementation of Multifractal Detrended Fluctuation Analysis called \texttt{MFDFA}.
MFDFA has found extensive application in the past two decades, yet a reliable, all-encompassing open-source software in \texttt{python} does not exist to this date.
In this library we have harnessed the most of \texttt{python}'s flexibility with handling matricial operations and multi-threaded polynomial fittings.
In this implementation we have included some of the more common extensions of MFDFA, including a simple empirical mode decomposition as a mechanism to detrend the data, a moving window to handle very short time series, and the extended Detrended Fluctuation Analysis, which can track a different scaling mechanism for non-stationary time series.
The \texttt{MFDFA} library can also handle missing values in the data with the aid of \texttt{numpy}'s \texttt{masked} time series.

We have initially turned to two classic numerically generated stochastic processes, fractional Ornstein--Uhlenbeck processes and Lévy-distributed motions, and uncovered their monofractal and multifractal with \texttt{MFDFA}.
Subsequently we have studied two real-world time series, the sunspot count from 1818 to 2020 and the quarter-hourly electricity price time series from 2015 to 2019.
For both we performed a multifractal analysis, unveiling their scaling properties and the strength of their multifractality.
We focused on \texttt{MFDFA}'s speed, the ability to handled missing data, and the integrated overlapping moving window.
The analysis displayed here covered only part of \texttt{MFDFA}'s integrated options, thus we leave the user to explore the other implemented methods, as the extended DFA and EMD detrending, as these are more specialised to particular research fields.

We hope with this contribution we open a door to fast MFDFA calculations that can the performed on a local machine without an extensive numerical effort and very long time runs, thus permitting in the future to analyse larger time series.

\section{Acknowledgements}
L.R.G. kindly thanks Francisco Meirinhos for all the help with \texttt{python}, and Fabian Harang, Marc Lagunas Merino, Anton Yurchenko-Tytarenko, Dennis Schroeder, Michele Giordano, Giulia di Nunno, and Fred Espen Benth for their support.
L.R.G and D.W gratefully acknowledge support by the Helmholtz Association, via the joint initiative \textit{Energy System 2050 - A Contribution of the Research Field Energy} and the grant \textit{Uncertainty Quantification -- From Data to Reliable Knowledge (UQ)}, with grant no.~ZT-I-0029, the scholarship funding from \textit{E.ON Stipendienfonds}, and the \textit{STORM - Stochastics for Time-Space Risk Models} project of the Research Council of Norway (RCN) no.~274410. 
This work was performed as part of the Helmholtz School for Data Science in Life, Earth and Energy (HDS-LEE).
J.K. was financed by the Ministry of Science and Higher Education of the Russian Federation within the framework of state support for the creation and development of World-Class Research Center ``Digital biodesign and personalised healthcare'', no.~075-15-2020-926.

\bibliographystyle{apsrev4-2}
\bibliography{bib}

\begin{thebibliography}{81}%
\makeatletter
\providecommand \@ifxundefined [1]{%
 \@ifx{#1\undefined}
}%
\providecommand \@ifnum [1]{%
 \ifnum #1\expandafter \@firstoftwo
 \else \expandafter \@secondoftwo
 \fi
}%
\providecommand \@ifx [1]{%
 \ifx #1\expandafter \@firstoftwo
 \else \expandafter \@secondoftwo
 \fi
}%
\providecommand \natexlab [1]{#1}%
\providecommand \enquote  [1]{``#1''}%
\providecommand \bibnamefont  [1]{#1}%
\providecommand \bibfnamefont [1]{#1}%
\providecommand \citenamefont [1]{#1}%
\providecommand \href@noop [0]{\@secondoftwo}%
\providecommand \href [0]{\begingroup \@sanitize@url \@href}%
\providecommand \@href[1]{\@@startlink{#1}\@@href}%
\providecommand \@@href[1]{\endgroup#1\@@endlink}%
\providecommand \@sanitize@url [0]{\catcode `\\12\catcode `\$12\catcode
  `\&12\catcode `\#12\catcode `\^12\catcode `\_12\catcode `\%12\relax}%
\providecommand \@@startlink[1]{}%
\providecommand \@@endlink[0]{}%
\providecommand \url  [0]{\begingroup\@sanitize@url \@url }%
\providecommand \@url [1]{\endgroup\@href {#1}{\urlprefix }}%
\providecommand \urlprefix  [0]{URL }%
\providecommand \Eprint [0]{\href }%
\providecommand \doibase [0]{https://doi.org/}%
\providecommand \selectlanguage [0]{\@gobble}%
\providecommand \bibinfo  [0]{\@secondoftwo}%
\providecommand \bibfield  [0]{\@secondoftwo}%
\providecommand \translation [1]{[#1]}%
\providecommand \BibitemOpen [0]{}%
\providecommand \bibitemStop [0]{}%
\providecommand \bibitemNoStop [0]{.\EOS\space}%
\providecommand \EOS [0]{\spacefactor3000\relax}%
\providecommand \BibitemShut  [1]{\csname bibitem#1\endcsname}%
\let\auto@bib@innerbib\@empty
\bibitem [{\citenamefont {Peng}\ \emph {et~al.}(1994)\citenamefont {Peng},
  \citenamefont {Buldyrev}, \citenamefont {Havlin}, \citenamefont {Simons},
  \citenamefont {Stanley},\ and\ \citenamefont {Goldberger}}]{Peng1994}%
  \BibitemOpen
  \bibfield  {author} {\bibinfo {author} {\bibfnamefont {C.-K.}\ \bibnamefont
  {Peng}}, \bibinfo {author} {\bibfnamefont {S.~V.}\ \bibnamefont {Buldyrev}},
  \bibinfo {author} {\bibfnamefont {S.}~\bibnamefont {Havlin}}, \bibinfo
  {author} {\bibfnamefont {M.}~\bibnamefont {Simons}}, \bibinfo {author}
  {\bibfnamefont {H.~E.}\ \bibnamefont {Stanley}},\ and\ \bibinfo {author}
  {\bibfnamefont {A.~L.}\ \bibnamefont {Goldberger}},\ }\href
  {https://doi.org/10.1103/PhysRevE.49.1685} {\bibfield  {journal} {\bibinfo
  {journal} {Physical Review E}\ }\textbf {\bibinfo {volume} {49}},\ \bibinfo
  {pages} {1685} (\bibinfo {year} {1994})}\BibitemShut {NoStop}%
\bibitem [{\citenamefont {Peng}\ \emph {et~al.}(1995)\citenamefont {Peng},
  \citenamefont {Havlin}, \citenamefont {Stanley},\ and\ \citenamefont
  {Goldberger}}]{Peng1995}%
  \BibitemOpen
  \bibfield  {author} {\bibinfo {author} {\bibfnamefont {C.}~\bibnamefont
  {Peng}}, \bibinfo {author} {\bibfnamefont {S.}~\bibnamefont {Havlin}},
  \bibinfo {author} {\bibfnamefont {H.~E.}\ \bibnamefont {Stanley}},\ and\
  \bibinfo {author} {\bibfnamefont {A.~L.}\ \bibnamefont {Goldberger}},\ }\href
  {https://doi.org/10.1063/1.166141} {\bibfield  {journal} {\bibinfo  {journal}
  {Chaos: An Interdisciplinary Journal of Nonlinear Science}\ }\textbf
  {\bibinfo {volume} {5}},\ \bibinfo {pages} {82} (\bibinfo {year}
  {1995})}\BibitemShut {NoStop}%
\bibitem [{\citenamefont {Kantelhardt}\ \emph {et~al.}(2002)\citenamefont
  {Kantelhardt}, \citenamefont {Zschiegner}, \citenamefont {Koscielny-Bunde},
  \citenamefont {Havlin}, \citenamefont {Bunde},\ and\ \citenamefont
  {Stanley}}]{Kantelhardt2002}%
  \BibitemOpen
  \bibfield  {author} {\bibinfo {author} {\bibfnamefont {J.~W.}\ \bibnamefont
  {Kantelhardt}}, \bibinfo {author} {\bibfnamefont {S.~A.}\ \bibnamefont
  {Zschiegner}}, \bibinfo {author} {\bibfnamefont {E.}~\bibnamefont
  {Koscielny-Bunde}}, \bibinfo {author} {\bibfnamefont {S.}~\bibnamefont
  {Havlin}}, \bibinfo {author} {\bibfnamefont {A.}~\bibnamefont {Bunde}},\ and\
  \bibinfo {author} {\bibfnamefont {H.}~\bibnamefont {Stanley}},\ }\href
  {https://doi.org/10.1016/S0378-4371(02)01383-3} {\bibfield  {journal}
  {\bibinfo  {journal} {Physica A: Statistical Mechanics and its Applications}\
  }\textbf {\bibinfo {volume} {316}},\ \bibinfo {pages} {87} (\bibinfo {year}
  {2002})}\BibitemShut {NoStop}%
\bibitem [{\citenamefont {Wu}\ and\ \citenamefont {Huang}(2004)}]{Zhaohua2004}%
  \BibitemOpen
  \bibfield  {author} {\bibinfo {author} {\bibfnamefont {Z.}~\bibnamefont
  {Wu}}\ and\ \bibinfo {author} {\bibfnamefont {N.~E.}\ \bibnamefont {Huang}},\
  }\href {https://doi.org/10.1098/rspa.2003.1221} {\bibfield  {journal}
  {\bibinfo  {journal} {Proceedings of the Royal Society of London. Series A:
  Mathematical, Physical and Engineering Sciences}\ }\textbf {\bibinfo {volume}
  {460}},\ \bibinfo {pages} {1597} (\bibinfo {year} {2004})}\BibitemShut
  {NoStop}%
\bibitem [{\citenamefont {Wu}\ and\ \citenamefont {Huang}(2009)}]{Zhaohua2009}%
  \BibitemOpen
  \bibfield  {author} {\bibinfo {author} {\bibfnamefont {Z.}~\bibnamefont
  {Wu}}\ and\ \bibinfo {author} {\bibfnamefont {N.~E.}\ \bibnamefont {Huang}},\
  }\href {https://doi.org/10.1142/S1793536909000047} {\bibfield  {journal}
  {\bibinfo  {journal} {Advances in Adaptive Data Analysis}\ }\textbf {\bibinfo
  {volume} {01}},\ \bibinfo {pages} {1} (\bibinfo {year} {2009})}\BibitemShut
  {NoStop}%
\bibitem [{\citenamefont {Qian}\ \emph {et~al.}(2011)\citenamefont {Qian},
  \citenamefont {Gu},\ and\ \citenamefont {Zhou}}]{Qian2011}%
  \BibitemOpen
  \bibfield  {author} {\bibinfo {author} {\bibfnamefont {X.-Y.}\ \bibnamefont
  {Qian}}, \bibinfo {author} {\bibfnamefont {G.-F.}\ \bibnamefont {Gu}},\ and\
  \bibinfo {author} {\bibfnamefont {W.-X.}\ \bibnamefont {Zhou}},\ }\href
  {https://doi.org/10.1016/j.physa.2011.07.008} {\bibfield  {journal} {\bibinfo
   {journal} {Physica A: Statistical Mechanics and its Applications}\ }\textbf
  {\bibinfo {volume} {390}},\ \bibinfo {pages} {4388} (\bibinfo {year}
  {2011})}\BibitemShut {NoStop}%
\bibitem [{\citenamefont {Zhang}\ \emph {et~al.}(2019)\citenamefont {Zhang},
  \citenamefont {Zhang}, \citenamefont {Qiu}, \citenamefont {Zhang},
  \citenamefont {Sun}, \citenamefont {Gui},\ and\ \citenamefont
  {Zhang}}]{Zhang2019}%
  \BibitemOpen
  \bibfield  {author} {\bibinfo {author} {\bibfnamefont {X.}~\bibnamefont
  {Zhang}}, \bibinfo {author} {\bibfnamefont {G.}~\bibnamefont {Zhang}},
  \bibinfo {author} {\bibfnamefont {L.}~\bibnamefont {Qiu}}, \bibinfo {author}
  {\bibfnamefont {B.}~\bibnamefont {Zhang}}, \bibinfo {author} {\bibfnamefont
  {Y.}~\bibnamefont {Sun}}, \bibinfo {author} {\bibfnamefont {Z.}~\bibnamefont
  {Gui}},\ and\ \bibinfo {author} {\bibfnamefont {Q.}~\bibnamefont {Zhang}},\
  }\href {https://doi.org/10.3390/w11050891} {\bibfield  {journal} {\bibinfo
  {journal} {Water}\ }\textbf {\bibinfo {volume} {11}},\ \bibinfo {pages} {891}
  (\bibinfo {year} {2019})}\BibitemShut {NoStop}%
\bibitem [{\citenamefont {Zhou}\ and\ \citenamefont {Leung}(2010)}]{Zhou2010}%
  \BibitemOpen
  \bibfield  {author} {\bibinfo {author} {\bibfnamefont {Y.}~\bibnamefont
  {Zhou}}\ and\ \bibinfo {author} {\bibfnamefont {Y.}~\bibnamefont {Leung}},\
  }\href {https://doi.org/10.1088/1742-5468/2010/06/p06021} {\bibfield
  {journal} {\bibinfo  {journal} {Journal of Statistical Mechanics: Theory and
  Experiment}\ }\textbf {\bibinfo {volume} {2010}},\ \bibinfo {pages} {P06021}
  (\bibinfo {year} {2010})}\BibitemShut {NoStop}%
\bibitem [{\citenamefont {Lai}\ \emph {et~al.}(2019)\citenamefont {Lai},
  \citenamefont {Wan},\ and\ \citenamefont {Zeng}}]{Lai2019}%
  \BibitemOpen
  \bibfield  {author} {\bibinfo {author} {\bibfnamefont {S.}~\bibnamefont
  {Lai}}, \bibinfo {author} {\bibfnamefont {L.}~\bibnamefont {Wan}},\ and\
  \bibinfo {author} {\bibfnamefont {X.}~\bibnamefont {Zeng}},\ }\href
  {https://doi.org/10.1088/1742-6596/1345/4/042086} {\bibfield  {journal}
  {\bibinfo  {journal} {Journal of Physics: Conference Series}\ }\textbf
  {\bibinfo {volume} {1345}},\ \bibinfo {pages} {042086} (\bibinfo {year}
  {2019})}\BibitemShut {NoStop}%
\bibitem [{\citenamefont {Pavlov}\ \emph
  {et~al.}(2020{\natexlab{a}})\citenamefont {Pavlov}, \citenamefont
  {Abdurashitov}, \citenamefont {Koronovskii}, \citenamefont {Pavlova},
  \citenamefont {Semyachkina-Glushkovskaya},\ and\ \citenamefont
  {Kurths}}]{Pavlov2020a}%
  \BibitemOpen
  \bibfield  {author} {\bibinfo {author} {\bibfnamefont {A.~N.}\ \bibnamefont
  {Pavlov}}, \bibinfo {author} {\bibfnamefont {A.~S.}\ \bibnamefont
  {Abdurashitov}}, \bibinfo {author} {\bibfnamefont {A.~A.}\ \bibnamefont
  {Koronovskii}}, \bibinfo {author} {\bibfnamefont {O.~N.}\ \bibnamefont
  {Pavlova}}, \bibinfo {author} {\bibfnamefont {O.~V.}\ \bibnamefont
  {Semyachkina-Glushkovskaya}},\ and\ \bibinfo {author} {\bibfnamefont
  {J.}~\bibnamefont {Kurths}},\ }\href
  {https://doi.org/10.1016/j.cnsns.2020.105232} {\bibfield  {journal} {\bibinfo
   {journal} {Communications in Nonlinear Science and Numerical Simulation}\
  }\textbf {\bibinfo {volume} {85}},\ \bibinfo {pages} {105232} (\bibinfo
  {year} {2020}{\natexlab{a}})}\BibitemShut {NoStop}%
\bibitem [{\citenamefont {Pavlov}\ \emph
  {et~al.}(2020{\natexlab{b}})\citenamefont {Pavlov}, \citenamefont
  {Dubrovsky}, \citenamefont {Koronovskii}, \citenamefont {Pavlova},
  \citenamefont {Semyachkina-Glushkovskaya},\ and\ \citenamefont
  {Kurths}}]{Pavlov2020b}%
  \BibitemOpen
  \bibfield  {author} {\bibinfo {author} {\bibfnamefont {A.~N.}\ \bibnamefont
  {Pavlov}}, \bibinfo {author} {\bibfnamefont {A.~I.}\ \bibnamefont
  {Dubrovsky}}, \bibinfo {author} {\bibfnamefont {A.~A.}\ \bibnamefont
  {Koronovskii}}, \bibinfo {author} {\bibfnamefont {O.~N.}\ \bibnamefont
  {Pavlova}}, \bibinfo {author} {\bibfnamefont {O.~V.}\ \bibnamefont
  {Semyachkina-Glushkovskaya}},\ and\ \bibinfo {author} {\bibfnamefont
  {J.}~\bibnamefont {Kurths}},\ }\href {https://doi.org/10.1063/5.0011823}
  {\bibfield  {journal} {\bibinfo  {journal} {Chaos: An Interdisciplinary
  Journal of Nonlinear Science}\ }\textbf {\bibinfo {volume} {30}},\ \bibinfo
  {pages} {073138} (\bibinfo {year} {2020}{\natexlab{b}})}\BibitemShut
  {NoStop}%
\bibitem [{\citenamefont {Pavlov}\ \emph
  {et~al.}(2020{\natexlab{c}})\citenamefont {Pavlov}, \citenamefont
  {Dubrovsky}, \citenamefont {{Koronovskii Jr}}, \citenamefont {Pavlova},
  \citenamefont {Semyachkina-Glushkovskaya},\ and\ \citenamefont
  {Kurths}}]{Pavlov2020c}%
  \BibitemOpen
  \bibfield  {author} {\bibinfo {author} {\bibfnamefont {A.}~\bibnamefont
  {Pavlov}}, \bibinfo {author} {\bibfnamefont {A.}~\bibnamefont {Dubrovsky}},
  \bibinfo {author} {\bibfnamefont {A.}~\bibnamefont {{Koronovskii Jr}}},
  \bibinfo {author} {\bibfnamefont {O.}~\bibnamefont {Pavlova}}, \bibinfo
  {author} {\bibfnamefont {O.}~\bibnamefont {Semyachkina-Glushkovskaya}},\ and\
  \bibinfo {author} {\bibfnamefont {J.}~\bibnamefont {Kurths}},\ }\href
  {https://doi.org/10.1016/j.chaos.2020.109989} {\bibfield  {journal} {\bibinfo
   {journal} {Chaos, Solitons \& Fractals}\ }\textbf {\bibinfo {volume}
  {139}},\ \bibinfo {pages} {109989} (\bibinfo {year}
  {2020}{\natexlab{c}})}\BibitemShut {NoStop}%
\bibitem [{\citenamefont {Pavlov}\ \emph {et~al.}(2021)\citenamefont {Pavlov},
  \citenamefont {Pavlova}, \citenamefont {Semyachkina-Glushkovskaya},\ and\
  \citenamefont {Kurths}}]{Pavlov2021}%
  \BibitemOpen
  \bibfield  {author} {\bibinfo {author} {\bibfnamefont {A.~N.}\ \bibnamefont
  {Pavlov}}, \bibinfo {author} {\bibfnamefont {O.~N.}\ \bibnamefont {Pavlova}},
  \bibinfo {author} {\bibfnamefont {O.~V.}\ \bibnamefont
  {Semyachkina-Glushkovskaya}},\ and\ \bibinfo {author} {\bibfnamefont
  {J.}~\bibnamefont {Kurths}},\ }\href
  {https://doi.org/10.1140/epjp/s13360-020-00980-x} {\bibfield  {journal}
  {\bibinfo  {journal} {The European Physical Journal Plus}\ }\textbf {\bibinfo
  {volume} {136}},\ \bibinfo {pages} {10} (\bibinfo {year} {2021})}\BibitemShut
  {NoStop}%
\bibitem [{\citenamefont {Podobnik}\ and\ \citenamefont
  {Stanley}(2008)}]{Podobnik2008}%
  \BibitemOpen
  \bibfield  {author} {\bibinfo {author} {\bibfnamefont {B.}~\bibnamefont
  {Podobnik}}\ and\ \bibinfo {author} {\bibfnamefont {H.~E.}\ \bibnamefont
  {Stanley}},\ }\href {https://doi.org/10.1103/PhysRevLett.100.084102}
  {\bibfield  {journal} {\bibinfo  {journal} {Physical Review Letters}\
  }\textbf {\bibinfo {volume} {100}},\ \bibinfo {pages} {084102} (\bibinfo
  {year} {2008})}\BibitemShut {NoStop}%
\bibitem [{\citenamefont {Zhou}(2008)}]{Zhou2008}%
  \BibitemOpen
  \bibfield  {author} {\bibinfo {author} {\bibfnamefont {W.-X.}\ \bibnamefont
  {Zhou}},\ }\href {https://doi.org/10.1103/PhysRevE.77.066211} {\bibfield
  {journal} {\bibinfo  {journal} {Physical Review E}\ }\textbf {\bibinfo
  {volume} {77}},\ \bibinfo {pages} {066211} (\bibinfo {year}
  {2008})}\BibitemShut {NoStop}%
\bibitem [{\citenamefont {Alvarez-Ramirez}\ \emph {et~al.}(2009)\citenamefont
  {Alvarez-Ramirez}, \citenamefont {Rodriguez},\ and\ \citenamefont
  {Echeverria}}]{Alvarez-Ramirez2009}%
  \BibitemOpen
  \bibfield  {author} {\bibinfo {author} {\bibfnamefont {J.}~\bibnamefont
  {Alvarez-Ramirez}}, \bibinfo {author} {\bibfnamefont {E.}~\bibnamefont
  {Rodriguez}},\ and\ \bibinfo {author} {\bibfnamefont {J.~C.}\ \bibnamefont
  {Echeverria}},\ }\href {https://doi.org/10.1103/PhysRevE.79.057202}
  {\bibfield  {journal} {\bibinfo  {journal} {Physical Review E}\ }\textbf
  {\bibinfo {volume} {79}},\ \bibinfo {pages} {057202} (\bibinfo {year}
  {2009})}\BibitemShut {NoStop}%
\bibitem [{\citenamefont {Chianca}\ \emph {et~al.}(2005)\citenamefont
  {Chianca}, \citenamefont {Ticona},\ and\ \citenamefont
  {Penna}}]{Chianca2005}%
  \BibitemOpen
  \bibfield  {author} {\bibinfo {author} {\bibfnamefont {C.}~\bibnamefont
  {Chianca}}, \bibinfo {author} {\bibfnamefont {A.}~\bibnamefont {Ticona}},\
  and\ \bibinfo {author} {\bibfnamefont {T.}~\bibnamefont {Penna}},\ }\href
  {https://doi.org/10.1016/j.physa.2005.03.047} {\bibfield  {journal} {\bibinfo
   {journal} {Physica A: Statistical Mechanics and its Applications}\ }\textbf
  {\bibinfo {volume} {357}},\ \bibinfo {pages} {447} (\bibinfo {year}
  {2005})}\BibitemShut {NoStop}%
\bibitem [{\citenamefont {Hu}\ \emph {et~al.}(2001)\citenamefont {Hu},
  \citenamefont {Ivanov}, \citenamefont {Chen}, \citenamefont {Carpena},\ and\
  \citenamefont {Eugene~Stanley}}]{Hu2001}%
  \BibitemOpen
  \bibfield  {author} {\bibinfo {author} {\bibfnamefont {K.}~\bibnamefont
  {Hu}}, \bibinfo {author} {\bibfnamefont {P.~C.}\ \bibnamefont {Ivanov}},
  \bibinfo {author} {\bibfnamefont {Z.}~\bibnamefont {Chen}}, \bibinfo {author}
  {\bibfnamefont {P.}~\bibnamefont {Carpena}},\ and\ \bibinfo {author}
  {\bibfnamefont {H.}~\bibnamefont {Eugene~Stanley}},\ }\href
  {https://doi.org/10.1103/PhysRevE.64.011114} {\bibfield  {journal} {\bibinfo
  {journal} {Physical Review E}\ }\textbf {\bibinfo {volume} {64}},\ \bibinfo
  {pages} {011114} (\bibinfo {year} {2001})}\BibitemShut {NoStop}%
\bibitem [{\citenamefont {Ivanov}\ \emph {et~al.}(1999)\citenamefont {Ivanov},
  \citenamefont {Amaral}, \citenamefont {Goldberger}, \citenamefont {Havlin},
  \citenamefont {Rosenblum}, \citenamefont {Struzik},\ and\ \citenamefont
  {Stanley}}]{Ivanov1999}%
  \BibitemOpen
  \bibfield  {author} {\bibinfo {author} {\bibfnamefont {P.~C.}\ \bibnamefont
  {Ivanov}}, \bibinfo {author} {\bibfnamefont {L.~A.~N.}\ \bibnamefont
  {Amaral}}, \bibinfo {author} {\bibfnamefont {A.~L.}\ \bibnamefont
  {Goldberger}}, \bibinfo {author} {\bibfnamefont {S.}~\bibnamefont {Havlin}},
  \bibinfo {author} {\bibfnamefont {M.~G.}\ \bibnamefont {Rosenblum}}, \bibinfo
  {author} {\bibfnamefont {Z.~R.}\ \bibnamefont {Struzik}},\ and\ \bibinfo
  {author} {\bibfnamefont {H.~E.}\ \bibnamefont {Stanley}},\ }\href
  {https://doi.org/10.1038/20924} {\bibfield  {journal} {\bibinfo  {journal}
  {Nature}\ }\textbf {\bibinfo {volume} {399}},\ \bibinfo {pages} {461}
  (\bibinfo {year} {1999})}\BibitemShut {NoStop}%
\bibitem [{\citenamefont {Dutta}\ \emph {et~al.}(2013)\citenamefont {Dutta},
  \citenamefont {Ghosh},\ and\ \citenamefont {Chatterjee}}]{Srimonti2013}%
  \BibitemOpen
  \bibfield  {author} {\bibinfo {author} {\bibfnamefont {S.}~\bibnamefont
  {Dutta}}, \bibinfo {author} {\bibfnamefont {D.}~\bibnamefont {Ghosh}},\ and\
  \bibinfo {author} {\bibfnamefont {S.}~\bibnamefont {Chatterjee}},\ }\href
  {https://doi.org/10.3389/fphys.2013.00274} {\bibfield  {journal} {\bibinfo
  {journal} {Frontiers in Physiology}\ }\textbf {\bibinfo {volume} {4}},\
  \bibinfo {pages} {274} (\bibinfo {year} {2013})}\BibitemShut {NoStop}%
\bibitem [{\citenamefont {Madanchi}\ \emph {et~al.}(2020)\citenamefont
  {Madanchi}, \citenamefont {Taghavi-Shahri}, \citenamefont {Taghavi-Shahri},\
  and\ \citenamefont {Rahimi~Tabar}}]{Madanchi2020}%
  \BibitemOpen
  \bibfield  {author} {\bibinfo {author} {\bibfnamefont {A.}~\bibnamefont
  {Madanchi}}, \bibinfo {author} {\bibfnamefont {F.}~\bibnamefont
  {Taghavi-Shahri}}, \bibinfo {author} {\bibfnamefont {S.~M.}\ \bibnamefont
  {Taghavi-Shahri}},\ and\ \bibinfo {author} {\bibfnamefont {M.~R.}\
  \bibnamefont {Rahimi~Tabar}},\ }\href
  {https://doi.org/10.1140/epjb/e2020-100561-4} {\bibfield  {journal} {\bibinfo
   {journal} {The European Physical Journal B}\ }\textbf {\bibinfo {volume}
  {93}},\ \bibinfo {pages} {126} (\bibinfo {year} {2020})}\BibitemShut
  {NoStop}%
\bibitem [{\citenamefont {Movahed}\ \emph {et~al.}(2011)\citenamefont
  {Movahed}, \citenamefont {Ghasemi}, \citenamefont {Rahvar},\ and\
  \citenamefont {Tabar}}]{Movahed2011}%
  \BibitemOpen
  \bibfield  {author} {\bibinfo {author} {\bibfnamefont {M.~S.}\ \bibnamefont
  {Movahed}}, \bibinfo {author} {\bibfnamefont {F.}~\bibnamefont {Ghasemi}},
  \bibinfo {author} {\bibfnamefont {S.}~\bibnamefont {Rahvar}},\ and\ \bibinfo
  {author} {\bibfnamefont {M.~R.~R.}\ \bibnamefont {Tabar}},\ }\href
  {https://doi.org/10.1103/PhysRevE.84.021103} {\bibfield  {journal} {\bibinfo
  {journal} {Physical Review E}\ }\textbf {\bibinfo {volume} {84}},\ \bibinfo
  {pages} {021103} (\bibinfo {year} {2011})}\BibitemShut {NoStop}%
\bibitem [{\citenamefont {Movahed}\ \emph {et~al.}(2013)\citenamefont
  {Movahed}, \citenamefont {Javanmardi},\ and\ \citenamefont
  {Sheth}}]{Movahed2013}%
  \BibitemOpen
  \bibfield  {author} {\bibinfo {author} {\bibfnamefont {M.~S.}\ \bibnamefont
  {Movahed}}, \bibinfo {author} {\bibfnamefont {B.}~\bibnamefont
  {Javanmardi}},\ and\ \bibinfo {author} {\bibfnamefont {R.~K.}\ \bibnamefont
  {Sheth}},\ }\href {https://doi.org/10.1093/mnras/stt1284} {\bibfield
  {journal} {\bibinfo  {journal} {Monthly Notices of the Royal Astronomical
  Society}\ }\textbf {\bibinfo {volume} {434}},\ \bibinfo {pages} {3597}
  (\bibinfo {year} {2013})}\BibitemShut {NoStop}%
\bibitem [{\citenamefont {Telesca}\ \emph {et~al.}(2005)\citenamefont
  {Telesca}, \citenamefont {Lapenna},\ and\ \citenamefont
  {Macchiato}}]{Telesca2005}%
  \BibitemOpen
  \bibfield  {author} {\bibinfo {author} {\bibfnamefont {L.}~\bibnamefont
  {Telesca}}, \bibinfo {author} {\bibfnamefont {V.}~\bibnamefont {Lapenna}},\
  and\ \bibinfo {author} {\bibfnamefont {M.}~\bibnamefont {Macchiato}},\ }\href
  {https://doi.org/10.1016/j.physa.2005.02.053} {\bibfield  {journal} {\bibinfo
   {journal} {Physica A: Statistical Mechanics and its Applications}\ }\textbf
  {\bibinfo {volume} {354}},\ \bibinfo {pages} {629} (\bibinfo {year}
  {2005})}\BibitemShut {NoStop}%
\bibitem [{\citenamefont {Shadkhoo}\ \emph {et~al.}(2009)\citenamefont
  {Shadkhoo}, \citenamefont {Ghanbarnejad}, \citenamefont {Jafari},\ and\
  \citenamefont {Tabar}}]{Shadkhoo2009}%
  \BibitemOpen
  \bibfield  {author} {\bibinfo {author} {\bibfnamefont {S.}~\bibnamefont
  {Shadkhoo}}, \bibinfo {author} {\bibfnamefont {F.}~\bibnamefont
  {Ghanbarnejad}}, \bibinfo {author} {\bibfnamefont {G.}~\bibnamefont
  {Jafari}},\ and\ \bibinfo {author} {\bibfnamefont {M.~R.~R.}\ \bibnamefont
  {Tabar}},\ }\href {https://doi.org/10.2478/s11534-009-0058-0} {\bibfield
  {journal} {\bibinfo  {journal} {Open Physics}\ }\textbf {\bibinfo {volume}
  {7}},\ \bibinfo {pages} {620} (\bibinfo {year} {2009})}\BibitemShut {NoStop}%
\bibitem [{\citenamefont {Sadegh~Movahed}\ \emph {et~al.}(2006)\citenamefont
  {Sadegh~Movahed}, \citenamefont {Jafari}, \citenamefont {Ghasemi},
  \citenamefont {Rahvar},\ and\ \citenamefont {Tabar}}]{Movahed2006}%
  \BibitemOpen
  \bibfield  {author} {\bibinfo {author} {\bibfnamefont {M.}~\bibnamefont
  {Sadegh~Movahed}}, \bibinfo {author} {\bibfnamefont {G.~R.}\ \bibnamefont
  {Jafari}}, \bibinfo {author} {\bibfnamefont {F.}~\bibnamefont {Ghasemi}},
  \bibinfo {author} {\bibfnamefont {S.}~\bibnamefont {Rahvar}},\ and\ \bibinfo
  {author} {\bibfnamefont {M.~R.~R.}\ \bibnamefont {Tabar}},\ }\href
  {https://doi.org/10.1088/1742-5468/2006/02/p02003} {\bibfield  {journal}
  {\bibinfo  {journal} {Journal of Statistical Mechanics: Theory and
  Experiment}\ }\textbf {\bibinfo {volume} {2006}},\ \bibinfo {pages} {P02003}
  (\bibinfo {year} {2006})}\BibitemShut {NoStop}%
\bibitem [{\citenamefont {Tanna}\ and\ \citenamefont
  {Pathak}(2014)}]{Tanna2014}%
  \BibitemOpen
  \bibfield  {author} {\bibinfo {author} {\bibfnamefont {H.~J.}\ \bibnamefont
  {Tanna}}\ and\ \bibinfo {author} {\bibfnamefont {K.~N.}\ \bibnamefont
  {Pathak}},\ }\href {https://doi.org/10.1007/s10509-013-1742-5} {\bibfield
  {journal} {\bibinfo  {journal} {Astrophysics and Space Science}\ }\textbf
  {\bibinfo {volume} {350}},\ \bibinfo {pages} {47} (\bibinfo {year}
  {2014})}\BibitemShut {NoStop}%
\bibitem [{\citenamefont {Meyer}\ and\ \citenamefont
  {Kantz}(2019)}]{Meyer2019}%
  \BibitemOpen
  \bibfield  {author} {\bibinfo {author} {\bibfnamefont {P.~G.}\ \bibnamefont
  {Meyer}}\ and\ \bibinfo {author} {\bibfnamefont {H.}~\bibnamefont {Kantz}},\
  }\href {https://doi.org/10.1007/s00382-019-04965-0} {\bibfield  {journal}
  {\bibinfo  {journal} {Climate Dynamics}\ }\textbf {\bibinfo {volume} {53}},\
  \bibinfo {pages} {6909} (\bibinfo {year} {2019})}\BibitemShut {NoStop}%
\bibitem [{\citenamefont {Pedron}(2010)}]{Pedron2010}%
  \BibitemOpen
  \bibfield  {author} {\bibinfo {author} {\bibfnamefont {I.~T.}\ \bibnamefont
  {Pedron}},\ }\href {https://doi.org/10.1088/1742-6596/246/1/012034}
  {\bibfield  {journal} {\bibinfo  {journal} {Journal of Physics: Conference
  Series}\ }\textbf {\bibinfo {volume} {246}},\ \bibinfo {pages} {012034}
  (\bibinfo {year} {2010})}\BibitemShut {NoStop}%
\bibitem [{\citenamefont {Tessier}\ \emph {et~al.}(1996)\citenamefont
  {Tessier}, \citenamefont {Lovejoy}, \citenamefont {Hubert}, \citenamefont
  {Schertzer},\ and\ \citenamefont {Pecknold}}]{Tessier1996}%
  \BibitemOpen
  \bibfield  {author} {\bibinfo {author} {\bibfnamefont {Y.}~\bibnamefont
  {Tessier}}, \bibinfo {author} {\bibfnamefont {S.}~\bibnamefont {Lovejoy}},
  \bibinfo {author} {\bibfnamefont {P.}~\bibnamefont {Hubert}}, \bibinfo
  {author} {\bibfnamefont {D.}~\bibnamefont {Schertzer}},\ and\ \bibinfo
  {author} {\bibfnamefont {S.}~\bibnamefont {Pecknold}},\ }\href
  {https://doi.org/10.1029/96JD01799} {\bibfield  {journal} {\bibinfo
  {journal} {Journal of Geophysical Research: Atmospheres}\ }\textbf {\bibinfo
  {volume} {101}},\ \bibinfo {pages} {26427} (\bibinfo {year}
  {1996})}\BibitemShut {NoStop}%
\bibitem [{\citenamefont {Matsoukas}\ \emph {et~al.}(2000)\citenamefont
  {Matsoukas}, \citenamefont {Islam},\ and\ \citenamefont
  {Rodriguez-Iturbe}}]{Matsoukas2000}%
  \BibitemOpen
  \bibfield  {author} {\bibinfo {author} {\bibfnamefont {C.}~\bibnamefont
  {Matsoukas}}, \bibinfo {author} {\bibfnamefont {S.}~\bibnamefont {Islam}},\
  and\ \bibinfo {author} {\bibfnamefont {I.}~\bibnamefont {Rodriguez-Iturbe}},\
  }\href {https://doi.org/10.1029/2000JD900419} {\bibfield  {journal} {\bibinfo
   {journal} {Journal of Geophysical Research: Atmospheres}\ }\textbf {\bibinfo
  {volume} {105}},\ \bibinfo {pages} {29165} (\bibinfo {year}
  {2000})}\BibitemShut {NoStop}%
\bibitem [{\citenamefont {Koutsoyiannis}(2003)}]{Koutsoyiannis2003}%
  \BibitemOpen
  \bibfield  {author} {\bibinfo {author} {\bibfnamefont {D.}~\bibnamefont
  {Koutsoyiannis}},\ }\href {https://doi.org/10.1623/hysj.48.1.3.43481}
  {\bibfield  {journal} {\bibinfo  {journal} {Hydrological Sciences Journal}\
  }\textbf {\bibinfo {volume} {48}},\ \bibinfo {pages} {3} (\bibinfo {year}
  {2003})}\BibitemShut {NoStop}%
\bibitem [{\citenamefont {Kantelhardt}\ \emph {et~al.}(2006)\citenamefont
  {Kantelhardt}, \citenamefont {Koscielny-Bunde}, \citenamefont {Rybski},
  \citenamefont {Braun}, \citenamefont {Bunde},\ and\ \citenamefont
  {Havlin}}]{Kantelhardt2006}%
  \BibitemOpen
  \bibfield  {author} {\bibinfo {author} {\bibfnamefont {J.~W.}\ \bibnamefont
  {Kantelhardt}}, \bibinfo {author} {\bibfnamefont {E.}~\bibnamefont
  {Koscielny-Bunde}}, \bibinfo {author} {\bibfnamefont {D.}~\bibnamefont
  {Rybski}}, \bibinfo {author} {\bibfnamefont {P.}~\bibnamefont {Braun}},
  \bibinfo {author} {\bibfnamefont {A.}~\bibnamefont {Bunde}},\ and\ \bibinfo
  {author} {\bibfnamefont {S.}~\bibnamefont {Havlin}},\ }\href
  {https://doi.org/10.1029/2005JD005881} {\bibfield  {journal} {\bibinfo
  {journal} {Journal of Geophysical Research: Atmospheres}\ }\textbf {\bibinfo
  {volume} {111}},\ \bibinfo {pages} {D01106} (\bibinfo {year}
  {2006})}\BibitemShut {NoStop}%
\bibitem [{\citenamefont {Zhang}\ \emph {et~al.}(2008)\citenamefont {Zhang},
  \citenamefont {Xu}, \citenamefont {Chen},\ and\ \citenamefont
  {Yu}}]{Zhang2008}%
  \BibitemOpen
  \bibfield  {author} {\bibinfo {author} {\bibfnamefont {Q.}~\bibnamefont
  {Zhang}}, \bibinfo {author} {\bibfnamefont {C.-Y.}\ \bibnamefont {Xu}},
  \bibinfo {author} {\bibfnamefont {Y.~D.}\ \bibnamefont {Chen}},\ and\
  \bibinfo {author} {\bibfnamefont {Z.}~\bibnamefont {Yu}},\ }\href
  {https://doi.org/10.1002/hyp.7119} {\bibfield  {journal} {\bibinfo  {journal}
  {Hydrological Processes}\ }\textbf {\bibinfo {volume} {22}},\ \bibinfo
  {pages} {4997} (\bibinfo {year} {2008})}\BibitemShut {NoStop}%
\bibitem [{\citenamefont {Rodr{\'i}guez}\ \emph {et~al.}(2013)\citenamefont
  {Rodr{\'i}guez}, \citenamefont {Casas},\ and\ \citenamefont
  {Reda{\~{n}}o}}]{Rodriguez2013}%
  \BibitemOpen
  \bibfield  {author} {\bibinfo {author} {\bibfnamefont {R.}~\bibnamefont
  {Rodr{\'i}guez}}, \bibinfo {author} {\bibfnamefont {M.~C.}\ \bibnamefont
  {Casas}},\ and\ \bibinfo {author} {\bibfnamefont {A.}~\bibnamefont
  {Reda{\~{n}}o}},\ }\href {https://doi.org/10.1007/s00703-013-0256-6}
  {\bibfield  {journal} {\bibinfo  {journal} {Meteorology and Atmospheric
  Physics}\ }\textbf {\bibinfo {volume} {121}},\ \bibinfo {pages} {181}
  (\bibinfo {year} {2013})}\BibitemShut {NoStop}%
\bibitem [{\citenamefont {Wu}\ \emph {et~al.}(2018)\citenamefont {Wu},
  \citenamefont {He}, \citenamefont {Wu}, \citenamefont {Lu}, \citenamefont
  {Gao},\ and\ \citenamefont {Xu}}]{Wu2018}%
  \BibitemOpen
  \bibfield  {author} {\bibinfo {author} {\bibfnamefont {Y.}~\bibnamefont
  {Wu}}, \bibinfo {author} {\bibfnamefont {Y.}~\bibnamefont {He}}, \bibinfo
  {author} {\bibfnamefont {M.}~\bibnamefont {Wu}}, \bibinfo {author}
  {\bibfnamefont {C.}~\bibnamefont {Lu}}, \bibinfo {author} {\bibfnamefont
  {S.}~\bibnamefont {Gao}},\ and\ \bibinfo {author} {\bibfnamefont
  {Y.}~\bibnamefont {Xu}},\ }\href {https://doi.org/10.1038/s41598-018-35032-z}
  {\bibfield  {journal} {\bibinfo  {journal} {Scientific Reports}\ }\textbf
  {\bibinfo {volume} {8}},\ \bibinfo {pages} {16553} (\bibinfo {year}
  {2018})}\BibitemShut {NoStop}%
\bibitem [{\citenamefont {Figueirêdo}\ \emph {et~al.}(2010)\citenamefont
  {Figueirêdo}, \citenamefont {Nogueira}, \citenamefont {Moret},\ and\
  \citenamefont {Coutinho}}]{Figueiredo2010}%
  \BibitemOpen
  \bibfield  {author} {\bibinfo {author} {\bibfnamefont {P.}~\bibnamefont
  {Figueirêdo}}, \bibinfo {author} {\bibfnamefont {E.}~\bibnamefont
  {Nogueira}}, \bibinfo {author} {\bibfnamefont {M.}~\bibnamefont {Moret}},\
  and\ \bibinfo {author} {\bibfnamefont {S.}~\bibnamefont {Coutinho}},\ }\href
  {https://doi.org/10.1016/j.physa.2009.11.045} {\bibfield  {journal} {\bibinfo
   {journal} {Physica A: Statistical Mechanics and its Applications}\ }\textbf
  {\bibinfo {volume} {389}},\ \bibinfo {pages} {2090} (\bibinfo {year}
  {2010})}\BibitemShut {NoStop}%
\bibitem [{\citenamefont {Zunino}\ \emph {et~al.}(2008)\citenamefont {Zunino},
  \citenamefont {Tabak}, \citenamefont {Figliola}, \citenamefont {Pérez},
  \citenamefont {Garavaglia},\ and\ \citenamefont {Rosso}}]{Zunino2008}%
  \BibitemOpen
  \bibfield  {author} {\bibinfo {author} {\bibfnamefont {L.}~\bibnamefont
  {Zunino}}, \bibinfo {author} {\bibfnamefont {B.}~\bibnamefont {Tabak}},
  \bibinfo {author} {\bibfnamefont {A.}~\bibnamefont {Figliola}}, \bibinfo
  {author} {\bibfnamefont {D.}~\bibnamefont {Pérez}}, \bibinfo {author}
  {\bibfnamefont {M.}~\bibnamefont {Garavaglia}},\ and\ \bibinfo {author}
  {\bibfnamefont {O.}~\bibnamefont {Rosso}},\ }\href
  {https://doi.org/10.1016/j.physa.2008.08.028} {\bibfield  {journal} {\bibinfo
   {journal} {Physica A: Statistical Mechanics and its Applications}\ }\textbf
  {\bibinfo {volume} {387}},\ \bibinfo {pages} {6558} (\bibinfo {year}
  {2008})}\BibitemShut {NoStop}%
\bibitem [{\citenamefont {Zunino}\ \emph {et~al.}(2009)\citenamefont {Zunino},
  \citenamefont {Figliola}, \citenamefont {Tabak}, \citenamefont {Pérez},
  \citenamefont {Garavaglia},\ and\ \citenamefont {Rosso}}]{Zunino2009}%
  \BibitemOpen
  \bibfield  {author} {\bibinfo {author} {\bibfnamefont {L.}~\bibnamefont
  {Zunino}}, \bibinfo {author} {\bibfnamefont {A.}~\bibnamefont {Figliola}},
  \bibinfo {author} {\bibfnamefont {B.~M.}\ \bibnamefont {Tabak}}, \bibinfo
  {author} {\bibfnamefont {D.~G.}\ \bibnamefont {Pérez}}, \bibinfo {author}
  {\bibfnamefont {M.}~\bibnamefont {Garavaglia}},\ and\ \bibinfo {author}
  {\bibfnamefont {O.~A.}\ \bibnamefont {Rosso}},\ }\href
  {https://doi.org/10.1016/j.chaos.2008.09.013} {\bibfield  {journal} {\bibinfo
   {journal} {Chaos, Solitons \& Fractals}\ }\textbf {\bibinfo {volume} {41}},\
  \bibinfo {pages} {2331} (\bibinfo {year} {2009})}\BibitemShut {NoStop}%
\bibitem [{\citenamefont {Grech}\ and\ \citenamefont
  {Pamu{\l}a}(2013)}]{Grech2013}%
  \BibitemOpen
  \bibfield  {author} {\bibinfo {author} {\bibfnamefont {D.}~\bibnamefont
  {Grech}}\ and\ \bibinfo {author} {\bibfnamefont {G.}~\bibnamefont
  {Pamu{\l}a}},\ }\href {https://doi.org/10.12693/aphyspola.123.529} {\bibfield
   {journal} {\bibinfo  {journal} {Acta Physica Polonica A}\ }\textbf {\bibinfo
  {volume} {123}},\ \bibinfo {pages} {529} (\bibinfo {year}
  {2013})}\BibitemShut {NoStop}%
\bibitem [{\citenamefont {Dro{\.{z}}d{\.{z}}}\ \emph
  {et~al.}(2018)\citenamefont {Dro{\.{z}}d{\.{z}}}, \citenamefont {Kowalski},
  \citenamefont {O{\'{s}}wi\c{e}cimka}, \citenamefont {Rak},\ and\
  \citenamefont {G\c{e}barowski}}]{Drozdz2018}%
  \BibitemOpen
  \bibfield  {author} {\bibinfo {author} {\bibfnamefont {S.}~\bibnamefont
  {Dro{\.{z}}d{\.{z}}}}, \bibinfo {author} {\bibfnamefont {R.}~\bibnamefont
  {Kowalski}}, \bibinfo {author} {\bibfnamefont {P.}~\bibnamefont
  {O{\'{s}}wi\c{e}cimka}}, \bibinfo {author} {\bibfnamefont {R.}~\bibnamefont
  {Rak}},\ and\ \bibinfo {author} {\bibfnamefont {R.}~\bibnamefont
  {G\c{e}barowski}},\ }\href {https://doi.org/10.1155/2018/7015721} {\bibfield
  {journal} {\bibinfo  {journal} {Complexity}\ }\textbf {\bibinfo {volume}
  {2018}},\ \bibinfo {pages} {7015721} (\bibinfo {year} {2018})}\BibitemShut
  {NoStop}%
\bibitem [{\citenamefont {Lee}\ \emph {et~al.}(2018)\citenamefont {Lee},
  \citenamefont {Song}, \citenamefont {Kim},\ and\ \citenamefont
  {Chang}}]{Minhyuk2018}%
  \BibitemOpen
  \bibfield  {author} {\bibinfo {author} {\bibfnamefont {M.}~\bibnamefont
  {Lee}}, \bibinfo {author} {\bibfnamefont {J.~W.}\ \bibnamefont {Song}},
  \bibinfo {author} {\bibfnamefont {S.}~\bibnamefont {Kim}},\ and\ \bibinfo
  {author} {\bibfnamefont {W.}~\bibnamefont {Chang}},\ }\href
  {https://doi.org/10.1016/j.physa.2018.08.030} {\bibfield  {journal} {\bibinfo
   {journal} {Physica A: Statistical Mechanics and its Applications}\ }\textbf
  {\bibinfo {volume} {512}},\ \bibinfo {pages} {1278} (\bibinfo {year}
  {2018})}\BibitemShut {NoStop}%
\bibitem [{\citenamefont {Weron}\ \emph
  {et~al.}(2004{\natexlab{a}})\citenamefont {Weron}, \citenamefont {Simonsen},\
  and\ \citenamefont {Wilman}}]{Weron2004b}%
  \BibitemOpen
  \bibfield  {author} {\bibinfo {author} {\bibfnamefont {R.}~\bibnamefont
  {Weron}}, \bibinfo {author} {\bibfnamefont {I.}~\bibnamefont {Simonsen}},\
  and\ \bibinfo {author} {\bibfnamefont {P.}~\bibnamefont {Wilman}},\ }in\
  \href {https://doi.org/10.1007/978-4-431-53947-6_25} {\emph {\bibinfo
  {booktitle} {The Application of Econophysics}}},\ \bibinfo {editor} {edited
  by\ \bibinfo {editor} {\bibfnamefont {H.}~\bibnamefont {Takayasu}}}\
  (\bibinfo  {publisher} {Springer Japan},\ \bibinfo {address} {Tokyo},\
  \bibinfo {year} {2004})\ pp.\ \bibinfo {pages} {182--191}\BibitemShut
  {NoStop}%
\bibitem [{\citenamefont {Wang}\ \emph {et~al.}(2013)\citenamefont {Wang},
  \citenamefont {Liao}, \citenamefont {Li}, \citenamefont {Li},\ and\
  \citenamefont {Zhou}}]{Wang2013}%
  \BibitemOpen
  \bibfield  {author} {\bibinfo {author} {\bibfnamefont {F.}~\bibnamefont
  {Wang}}, \bibinfo {author} {\bibfnamefont {G.}~\bibnamefont {Liao}}, \bibinfo
  {author} {\bibfnamefont {J.}~\bibnamefont {Li}}, \bibinfo {author}
  {\bibfnamefont {X.}~\bibnamefont {Li}},\ and\ \bibinfo {author}
  {\bibfnamefont {T.}~\bibnamefont {Zhou}},\ }\href
  {https://doi.org/10.1016/j.physa.2013.07.039} {\bibfield  {journal} {\bibinfo
   {journal} {Physica A: Statistical Mechanics and its Applications}\ }\textbf
  {\bibinfo {volume} {392}},\ \bibinfo {pages} {5723} (\bibinfo {year}
  {2013})}\BibitemShut {NoStop}%
\bibitem [{\citenamefont {{Shalalfeh}}\ \emph {et~al.}(2016)\citenamefont
  {{Shalalfeh}}, \citenamefont {{Bogdan}},\ and\ \citenamefont
  {{Jonckheere}}}]{Shalalfeh2016}%
  \BibitemOpen
  \bibfield  {author} {\bibinfo {author} {\bibfnamefont {L.}~\bibnamefont
  {{Shalalfeh}}}, \bibinfo {author} {\bibfnamefont {P.}~\bibnamefont
  {{Bogdan}}},\ and\ \bibinfo {author} {\bibfnamefont {E.}~\bibnamefont
  {{Jonckheere}}},\ }in\ \href
  {https://doi.org/10.1109/SmartGridComm.2016.7778805} {\emph {\bibinfo
  {booktitle} {2016 {IEEE} International Conference on Smart Grid
  Communications (SmartGridComm)}}}\ (\bibinfo {year} {2016})\ pp.\ \bibinfo
  {pages} {466--471}\BibitemShut {NoStop}%
\bibitem [{\citenamefont {Rydin~Gorj{\~a}o}\ \emph {et~al.}(2020)\citenamefont
  {Rydin~Gorj{\~a}o}, \citenamefont {Jumar}, \citenamefont {Maass},
  \citenamefont {Hagenmeyer}, \citenamefont {Yalcin}, \citenamefont {Kruse},
  \citenamefont {Timme}, \citenamefont {Beck}, \citenamefont {Witthaut},\ and\
  \citenamefont {Sch{\"a}fer}}]{RydinGorjao2020}%
  \BibitemOpen
  \bibfield  {author} {\bibinfo {author} {\bibfnamefont {L.}~\bibnamefont
  {Rydin~Gorj{\~a}o}}, \bibinfo {author} {\bibfnamefont {R.}~\bibnamefont
  {Jumar}}, \bibinfo {author} {\bibfnamefont {H.}~\bibnamefont {Maass}},
  \bibinfo {author} {\bibfnamefont {V.}~\bibnamefont {Hagenmeyer}}, \bibinfo
  {author} {\bibfnamefont {G.~C.}\ \bibnamefont {Yalcin}}, \bibinfo {author}
  {\bibfnamefont {J.}~\bibnamefont {Kruse}}, \bibinfo {author} {\bibfnamefont
  {M.}~\bibnamefont {Timme}}, \bibinfo {author} {\bibfnamefont
  {C.}~\bibnamefont {Beck}}, \bibinfo {author} {\bibfnamefont {D.}~\bibnamefont
  {Witthaut}},\ and\ \bibinfo {author} {\bibfnamefont {B.}~\bibnamefont
  {Sch{\"a}fer}},\ }\href {https://doi.org/10.1038/s41467-020-19732-7}
  {\bibfield  {journal} {\bibinfo  {journal} {Nature Communications}\ }\textbf
  {\bibinfo {volume} {11}},\ \bibinfo {pages} {6362} (\bibinfo {year}
  {2020})}\BibitemShut {NoStop}%
\bibitem [{\citenamefont {Leung}\ \emph {et~al.}(2011)\citenamefont {Leung},
  \citenamefont {Ge},\ and\ \citenamefont {Yu}}]{Leung2011}%
  \BibitemOpen
  \bibfield  {author} {\bibinfo {author} {\bibfnamefont {Y.}~\bibnamefont
  {Leung}}, \bibinfo {author} {\bibfnamefont {E.}~\bibnamefont {Ge}},\ and\
  \bibinfo {author} {\bibfnamefont {Z.}~\bibnamefont {Yu}},\ }\href
  {https://doi.org/10.1080/00045608.2011.592733} {\bibfield  {journal}
  {\bibinfo  {journal} {Annals of the Association of American Geographers}\
  }\textbf {\bibinfo {volume} {101}},\ \bibinfo {pages} {1221} (\bibinfo {year}
  {2011})}\BibitemShut {NoStop}%
\bibitem [{\citenamefont {Jafari}\ \emph {et~al.}(2007)\citenamefont {Jafari},
  \citenamefont {Pedram},\ and\ \citenamefont {Hedayatifar}}]{Jafari2007}%
  \BibitemOpen
  \bibfield  {author} {\bibinfo {author} {\bibfnamefont {G.~R.}\ \bibnamefont
  {Jafari}}, \bibinfo {author} {\bibfnamefont {P.}~\bibnamefont {Pedram}},\
  and\ \bibinfo {author} {\bibfnamefont {L.}~\bibnamefont {Hedayatifar}},\
  }\href {https://doi.org/10.1088/1742-5468/2007/04/p04012} {\bibfield
  {journal} {\bibinfo  {journal} {Journal of Statistical Mechanics: Theory and
  Experiment}\ }\textbf {\bibinfo {volume} {2007}},\ \bibinfo {pages} {P04012}
  (\bibinfo {year} {2007})}\BibitemShut {NoStop}%
\bibitem [{\citenamefont {Telesca}\ and\ \citenamefont
  {Lovallo}(2011)}]{Telesca2011}%
  \BibitemOpen
  \bibfield  {author} {\bibinfo {author} {\bibfnamefont {L.}~\bibnamefont
  {Telesca}}\ and\ \bibinfo {author} {\bibfnamefont {M.}~\bibnamefont
  {Lovallo}},\ }\href {https://doi.org/10.1098/rspa.2011.0118} {\bibfield
  {journal} {\bibinfo  {journal} {Proceedings of the Royal Society A:
  Mathematical, Physical and Engineering Sciences}\ }\textbf {\bibinfo {volume}
  {467}},\ \bibinfo {pages} {3022} (\bibinfo {year} {2011})}\BibitemShut
  {NoStop}%
\bibitem [{\citenamefont {Ribeiro}\ \emph {et~al.}(2012)\citenamefont
  {Ribeiro}, \citenamefont {Zunino}, \citenamefont {Mendes},\ and\
  \citenamefont {Lenzi}}]{Ribeiro2012}%
  \BibitemOpen
  \bibfield  {author} {\bibinfo {author} {\bibfnamefont {H.~V.}\ \bibnamefont
  {Ribeiro}}, \bibinfo {author} {\bibfnamefont {L.}~\bibnamefont {Zunino}},
  \bibinfo {author} {\bibfnamefont {R.~S.}\ \bibnamefont {Mendes}},\ and\
  \bibinfo {author} {\bibfnamefont {E.~K.}\ \bibnamefont {Lenzi}},\ }\href
  {https://doi.org/10.1016/j.physa.2011.12.009} {\bibfield  {journal} {\bibinfo
   {journal} {Physica A: Statistical Mechanics and its Applications}\ }\textbf
  {\bibinfo {volume} {391}},\ \bibinfo {pages} {2421} (\bibinfo {year}
  {2012})}\BibitemShut {NoStop}%
\bibitem [{\citenamefont {Alados}\ and\ \citenamefont
  {Huffman}(2000)}]{Alados2000}%
  \BibitemOpen
  \bibfield  {author} {\bibinfo {author} {\bibfnamefont {C.~L.}\ \bibnamefont
  {Alados}}\ and\ \bibinfo {author} {\bibfnamefont {M.~A.}\ \bibnamefont
  {Huffman}},\ }\href {https://doi.org/10.1046/j.1439-0310.2000.00497.x}
  {\bibfield  {journal} {\bibinfo  {journal} {Ethology}\ }\textbf {\bibinfo
  {volume} {106}},\ \bibinfo {pages} {105} (\bibinfo {year}
  {2000})}\BibitemShut {NoStop}%
\bibitem [{\citenamefont {Rutherford}\ \emph {et~al.}(2003)\citenamefont
  {Rutherford}, \citenamefont {Haskell}, \citenamefont {Glasbey}, \citenamefont
  {Jones},\ and\ \citenamefont {Lawrence}}]{Rutherford2003}%
  \BibitemOpen
  \bibfield  {author} {\bibinfo {author} {\bibfnamefont {K.~M.}\ \bibnamefont
  {Rutherford}}, \bibinfo {author} {\bibfnamefont {M.~J.}\ \bibnamefont
  {Haskell}}, \bibinfo {author} {\bibfnamefont {C.}~\bibnamefont {Glasbey}},
  \bibinfo {author} {\bibfnamefont {R.}~\bibnamefont {Jones}},\ and\ \bibinfo
  {author} {\bibfnamefont {A.~B.}\ \bibnamefont {Lawrence}},\ }\href
  {https://doi.org/10.1016/S0168-1591(03)00115-1} {\bibfield  {journal}
  {\bibinfo  {journal} {Applied Animal Behaviour Science}\ }\textbf {\bibinfo
  {volume} {83}},\ \bibinfo {pages} {125} (\bibinfo {year} {2003})}\BibitemShut
  {NoStop}%
\bibitem [{\citenamefont {Li}\ \emph {et~al.}(2019)\citenamefont {Li},
  \citenamefont {Ma}, \citenamefont {Zhao},\ and\ \citenamefont
  {Cheng}}]{Li2019}%
  \BibitemOpen
  \bibfield  {author} {\bibinfo {author} {\bibfnamefont {J.}~\bibnamefont
  {Li}}, \bibinfo {author} {\bibfnamefont {X.}~\bibnamefont {Ma}}, \bibinfo
  {author} {\bibfnamefont {M.}~\bibnamefont {Zhao}},\ and\ \bibinfo {author}
  {\bibfnamefont {X.}~\bibnamefont {Cheng}},\ }\href
  {https://doi.org/10.3390/electronics8020209} {\bibfield  {journal} {\bibinfo
  {journal} {Electronics}\ }\textbf {\bibinfo {volume} {8}},\ \bibinfo {pages}
  {209} (\bibinfo {year} {2019})}\BibitemShut {NoStop}%
\bibitem [{\citenamefont {Madanchi}\ \emph {et~al.}(2021)\citenamefont
  {Madanchi}, \citenamefont {Yu}, \citenamefont {Tabar}, \citenamefont {Lee},\
  and\ \citenamefont {Rahbari}}]{Madanchi2021}%
  \BibitemOpen
  \bibfield  {author} {\bibinfo {author} {\bibfnamefont {A.}~\bibnamefont
  {Madanchi}}, \bibinfo {author} {\bibfnamefont {J.~W.}\ \bibnamefont {Yu}},
  \bibinfo {author} {\bibfnamefont {M.~R.~R.}\ \bibnamefont {Tabar}}, \bibinfo
  {author} {\bibfnamefont {W.~B.}\ \bibnamefont {Lee}},\ and\ \bibinfo {author}
  {\bibfnamefont {S.~E.~E.}\ \bibnamefont {Rahbari}},\ }\href
  {https://doi.org/10.1039/D0SM02039G} {\bibfield  {journal} {\bibinfo
  {journal} {Soft Matter}\ ,\ \bibinfo {pages} {\textit{Accepted Publication}}}
  (\bibinfo {year} {2021})}\BibitemShut {NoStop}%
\bibitem [{\citenamefont {Ihlen}(2012)}]{Ihlen2012}%
  \BibitemOpen
  \bibfield  {author} {\bibinfo {author} {\bibfnamefont {E.}~\bibnamefont
  {Ihlen}},\ }\href {https://doi.org/10.3389/fphys.2012.00141} {\bibfield
  {journal} {\bibinfo  {journal} {Frontiers in Physiology}\ }\textbf {\bibinfo
  {volume} {3}},\ \bibinfo {pages} {141} (\bibinfo {year} {2012})}\BibitemShut
  {NoStop}%
\bibitem [{\citenamefont {Laib}\ \emph
  {et~al.}(2018{\natexlab{a}})\citenamefont {Laib}, \citenamefont {Telesca},\
  and\ \citenamefont {Kanevski}}]{Laib2018a}%
  \BibitemOpen
  \bibfield  {author} {\bibinfo {author} {\bibfnamefont {M.}~\bibnamefont
  {Laib}}, \bibinfo {author} {\bibfnamefont {L.}~\bibnamefont {Telesca}},\ and\
  \bibinfo {author} {\bibfnamefont {M.}~\bibnamefont {Kanevski}},\ }\href
  {https://doi.org/10.1063/1.5022737} {\bibfield  {journal} {\bibinfo
  {journal} {Chaos: An Interdisciplinary Journal of Nonlinear Science}\
  }\textbf {\bibinfo {volume} {28}},\ \bibinfo {pages} {033108} (\bibinfo
  {year} {2018}{\natexlab{a}})}\BibitemShut {NoStop}%
\bibitem [{\citenamefont {Laib}\ \emph
  {et~al.}(2018{\natexlab{b}})\citenamefont {Laib}, \citenamefont {Golay},
  \citenamefont {Telesca},\ and\ \citenamefont {Kanevski}}]{Laib2018b}%
  \BibitemOpen
  \bibfield  {author} {\bibinfo {author} {\bibfnamefont {M.}~\bibnamefont
  {Laib}}, \bibinfo {author} {\bibfnamefont {J.}~\bibnamefont {Golay}},
  \bibinfo {author} {\bibfnamefont {L.}~\bibnamefont {Telesca}},\ and\ \bibinfo
  {author} {\bibfnamefont {M.}~\bibnamefont {Kanevski}},\ }\href
  {https://doi.org/10.1016/j.chaos.2018.02.024} {\bibfield  {journal} {\bibinfo
   {journal} {Chaos, Solitons \& Fractals}\ }\textbf {\bibinfo {volume}
  {109}},\ \bibinfo {pages} {118} (\bibinfo {year}
  {2018}{\natexlab{b}})}\BibitemShut {NoStop}%
\bibitem [{\citenamefont {Harris}\ \emph {et~al.}(2020)\citenamefont {Harris},
  \citenamefont {Millman}, \citenamefont {van~der Walt}, \citenamefont
  {Gommers}, \citenamefont {Virtanen}, \citenamefont {Cournapeau},
  \citenamefont {Wieser}, \citenamefont {Taylor}, \citenamefont {Berg},
  \citenamefont {Smith}, \citenamefont {Kern}, \citenamefont {Picus},
  \citenamefont {Hoyer}, \citenamefont {van Kerkwijk}, \citenamefont {Brett},
  \citenamefont {Haldane}, \citenamefont {del R{'{\i}}o}, \citenamefont
  {Wiebe}, \citenamefont {Peterson}, \citenamefont {G{'{e}}rard-Marchant},
  \citenamefont {Sheppard}, \citenamefont {Reddy}, \citenamefont {Weckesser},
  \citenamefont {Abbasi}, \citenamefont {Gohlke},\ and\ \citenamefont
  {Oliphant}}]{NumPy}%
  \BibitemOpen
  \bibfield  {author} {\bibinfo {author} {\bibfnamefont {C.~R.}\ \bibnamefont
  {Harris}}, \bibinfo {author} {\bibfnamefont {K.~J.}\ \bibnamefont {Millman}},
  \bibinfo {author} {\bibfnamefont {S.~J.}\ \bibnamefont {van~der Walt}},
  \bibinfo {author} {\bibfnamefont {R.}~\bibnamefont {Gommers}}, \bibinfo
  {author} {\bibfnamefont {P.}~\bibnamefont {Virtanen}}, \bibinfo {author}
  {\bibfnamefont {D.}~\bibnamefont {Cournapeau}}, \bibinfo {author}
  {\bibfnamefont {E.}~\bibnamefont {Wieser}}, \bibinfo {author} {\bibfnamefont
  {J.}~\bibnamefont {Taylor}}, \bibinfo {author} {\bibfnamefont
  {S.}~\bibnamefont {Berg}}, \bibinfo {author} {\bibfnamefont {N.~J.}\
  \bibnamefont {Smith}}, \bibinfo {author} {\bibfnamefont {R.}~\bibnamefont
  {Kern}}, \bibinfo {author} {\bibfnamefont {M.}~\bibnamefont {Picus}},
  \bibinfo {author} {\bibfnamefont {S.}~\bibnamefont {Hoyer}}, \bibinfo
  {author} {\bibfnamefont {M.~H.}\ \bibnamefont {van Kerkwijk}}, \bibinfo
  {author} {\bibfnamefont {M.}~\bibnamefont {Brett}}, \bibinfo {author}
  {\bibfnamefont {A.}~\bibnamefont {Haldane}}, \bibinfo {author} {\bibfnamefont
  {J.~F.}\ \bibnamefont {del R{'{\i}}o}}, \bibinfo {author} {\bibfnamefont
  {M.}~\bibnamefont {Wiebe}}, \bibinfo {author} {\bibfnamefont
  {P.}~\bibnamefont {Peterson}}, \bibinfo {author} {\bibfnamefont
  {P.}~\bibnamefont {G{'{e}}rard-Marchant}}, \bibinfo {author} {\bibfnamefont
  {K.}~\bibnamefont {Sheppard}}, \bibinfo {author} {\bibfnamefont
  {T.}~\bibnamefont {Reddy}}, \bibinfo {author} {\bibfnamefont
  {W.}~\bibnamefont {Weckesser}}, \bibinfo {author} {\bibfnamefont
  {H.}~\bibnamefont {Abbasi}}, \bibinfo {author} {\bibfnamefont
  {C.}~\bibnamefont {Gohlke}},\ and\ \bibinfo {author} {\bibfnamefont {T.~E.}\
  \bibnamefont {Oliphant}},\ }\href {https://doi.org/10.1038/s41586-020-2649-2}
  {\bibfield  {journal} {\bibinfo  {journal} {Nature}\ }\textbf {\bibinfo
  {volume} {585}},\ \bibinfo {pages} {357} (\bibinfo {year}
  {2020})}\BibitemShut {NoStop}%
\bibitem [{\citenamefont {Hurst}(1951)}]{Hurst1951}%
  \BibitemOpen
  \bibfield  {author} {\bibinfo {author} {\bibfnamefont {H.~E.}\ \bibnamefont
  {Hurst}},\ }\href {https://doi.org/10.1061/TACEAT.0006518} {\bibfield
  {journal} {\bibinfo  {journal} {Transactions of the American Society of Civil
  Engineers}\ }\textbf {\bibinfo {volume} {116}},\ \bibinfo {pages} {770}
  (\bibinfo {year} {1951})}\BibitemShut {NoStop}%
\bibitem [{\citenamefont {Hentschel}\ and\ \citenamefont
  {Procaccia}(1983)}]{Hentschel1983}%
  \BibitemOpen
  \bibfield  {author} {\bibinfo {author} {\bibfnamefont {H.}~\bibnamefont
  {Hentschel}}\ and\ \bibinfo {author} {\bibfnamefont {I.}~\bibnamefont
  {Procaccia}},\ }\href {https://doi.org/10.1016/0167-2789(83)90235-X}
  {\bibfield  {journal} {\bibinfo  {journal} {Physica D: Nonlinear Phenomena}\
  }\textbf {\bibinfo {volume} {8}},\ \bibinfo {pages} {435} (\bibinfo {year}
  {1983})}\BibitemShut {NoStop}%
\bibitem [{\citenamefont {Halsey}\ \emph {et~al.}(1986)\citenamefont {Halsey},
  \citenamefont {Jensen}, \citenamefont {Kadanoff}, \citenamefont {Procaccia},\
  and\ \citenamefont {Shraiman}}]{Halsey1986}%
  \BibitemOpen
  \bibfield  {author} {\bibinfo {author} {\bibfnamefont {T.~C.}\ \bibnamefont
  {Halsey}}, \bibinfo {author} {\bibfnamefont {M.~H.}\ \bibnamefont {Jensen}},
  \bibinfo {author} {\bibfnamefont {L.~P.}\ \bibnamefont {Kadanoff}}, \bibinfo
  {author} {\bibfnamefont {I.}~\bibnamefont {Procaccia}},\ and\ \bibinfo
  {author} {\bibfnamefont {B.~I.}\ \bibnamefont {Shraiman}},\ }\href
  {https://doi.org/10.1103/PhysRevA.33.1141} {\bibfield  {journal} {\bibinfo
  {journal} {Physical Review A}\ }\textbf {\bibinfo {volume} {33}},\ \bibinfo
  {pages} {1141} (\bibinfo {year} {1986})}\BibitemShut {NoStop}%
\bibitem [{\citenamefont {Kurths}\ and\ \citenamefont
  {Herzel}(1987)}]{Kurths1987}%
  \BibitemOpen
  \bibfield  {author} {\bibinfo {author} {\bibfnamefont {J.}~\bibnamefont
  {Kurths}}\ and\ \bibinfo {author} {\bibfnamefont {H.}~\bibnamefont
  {Herzel}},\ }\href {https://doi.org/10.1016/0167-2789(87)90099-6} {\bibfield
  {journal} {\bibinfo  {journal} {Physica D: Nonlinear Phenomena}\ }\textbf
  {\bibinfo {volume} {25}},\ \bibinfo {pages} {165} (\bibinfo {year}
  {1987})}\BibitemShut {NoStop}%
\bibitem [{\citenamefont {Meneveau}\ and\ \citenamefont
  {Sreenivasan}(1987)}]{Meneveau1987}%
  \BibitemOpen
  \bibfield  {author} {\bibinfo {author} {\bibfnamefont {C.}~\bibnamefont
  {Meneveau}}\ and\ \bibinfo {author} {\bibfnamefont {K.}~\bibnamefont
  {Sreenivasan}},\ }\href {https://doi.org/10.1016/0920-5632(87)90008-9}
  {\bibfield  {journal} {\bibinfo  {journal} {Nuclear Physics B - Proceedings
  Supplements}\ }\textbf {\bibinfo {volume} {2}},\ \bibinfo {pages} {49}
  (\bibinfo {year} {1987})}\BibitemShut {NoStop}%
\bibitem [{\citenamefont {Salat}\ \emph {et~al.}(2017)\citenamefont {Salat},
  \citenamefont {Murcio},\ and\ \citenamefont {Arcaute}}]{Salat2017}%
  \BibitemOpen
  \bibfield  {author} {\bibinfo {author} {\bibfnamefont {H.}~\bibnamefont
  {Salat}}, \bibinfo {author} {\bibfnamefont {R.}~\bibnamefont {Murcio}},\ and\
  \bibinfo {author} {\bibfnamefont {E.}~\bibnamefont {Arcaute}},\ }\href
  {https://doi.org/10.1016/j.physa.2017.01.041} {\bibfield  {journal} {\bibinfo
   {journal} {Physica A: Statistical Mechanics and its Applications}\ }\textbf
  {\bibinfo {volume} {473}},\ \bibinfo {pages} {467} (\bibinfo {year}
  {2017})}\BibitemShut {NoStop}%
\bibitem [{\citenamefont {Falconer}(2014)}]{Falconer2004}%
  \BibitemOpen
  \bibfield  {author} {\bibinfo {author} {\bibfnamefont {K.}~\bibnamefont
  {Falconer}},\ }\href@noop {} {\emph {\bibinfo {title} {Fractal geometry:
  mathematical foundations and applications}}},\ \bibinfo {edition} {3rd}\ ed.\
  (\bibinfo  {publisher} {John Wiley \& Sons, Chichester},\ \bibinfo {year}
  {2014})\BibitemShut {NoStop}%
\bibitem [{\citenamefont {Barab\'asi}\ and\ \citenamefont
  {Vicsek}(1991)}]{Barabasi1991}%
  \BibitemOpen
  \bibfield  {author} {\bibinfo {author} {\bibfnamefont {A.-L.}\ \bibnamefont
  {Barab\'asi}}\ and\ \bibinfo {author} {\bibfnamefont {T.}~\bibnamefont
  {Vicsek}},\ }\href {https://doi.org/10.1103/PhysRevA.44.2730} {\bibfield
  {journal} {\bibinfo  {journal} {Physical Review A}\ }\textbf {\bibinfo
  {volume} {44}},\ \bibinfo {pages} {2730} (\bibinfo {year}
  {1991})}\BibitemShut {NoStop}%
\bibitem [{\citenamefont {Serinaldi}(2010)}]{Serinaldi2010}%
  \BibitemOpen
  \bibfield  {author} {\bibinfo {author} {\bibfnamefont {F.}~\bibnamefont
  {Serinaldi}},\ }\href {https://doi.org/10.1016/j.physa.2010.02.044}
  {\bibfield  {journal} {\bibinfo  {journal} {Physica A: Statistical Mechanics
  and its Applications}\ }\textbf {\bibinfo {volume} {389}},\ \bibinfo {pages}
  {2770} (\bibinfo {year} {2010})}\BibitemShut {NoStop}%
\bibitem [{\citenamefont {Huang}\ \emph {et~al.}(1998)\citenamefont {Huang},
  \citenamefont {Shen}, \citenamefont {Long}, \citenamefont {Wu}, \citenamefont
  {Shih}, \citenamefont {Zheng}, \citenamefont {Yen}, \citenamefont {Tung},\
  and\ \citenamefont {Liu}}]{Huang1998}%
  \BibitemOpen
  \bibfield  {author} {\bibinfo {author} {\bibfnamefont {N.~E.}\ \bibnamefont
  {Huang}}, \bibinfo {author} {\bibfnamefont {Z.}~\bibnamefont {Shen}},
  \bibinfo {author} {\bibfnamefont {S.~R.}\ \bibnamefont {Long}}, \bibinfo
  {author} {\bibfnamefont {M.~C.}\ \bibnamefont {Wu}}, \bibinfo {author}
  {\bibfnamefont {H.~H.}\ \bibnamefont {Shih}}, \bibinfo {author}
  {\bibfnamefont {Q.}~\bibnamefont {Zheng}}, \bibinfo {author} {\bibfnamefont
  {N.-C.}\ \bibnamefont {Yen}}, \bibinfo {author} {\bibfnamefont {C.~C.}\
  \bibnamefont {Tung}},\ and\ \bibinfo {author} {\bibfnamefont {H.~H.}\
  \bibnamefont {Liu}},\ }\href {https://doi.org/10.1098/rspa.1998.0193}
  {\bibfield  {journal} {\bibinfo  {journal} {Proceedings of the Royal Society
  of London. Series A: Mathematical, Physical and Engineering Sciences}\
  }\textbf {\bibinfo {volume} {454}},\ \bibinfo {pages} {903} (\bibinfo {year}
  {1998})}\BibitemShut {NoStop}%
\bibitem [{\citenamefont {Tabar}(2019)}]{Tabar2019}%
  \BibitemOpen
  \bibfield  {author} {\bibinfo {author} {\bibfnamefont {M.~R.~R.}\
  \bibnamefont {Tabar}},\ }\href {https://doi.org/10.1007/978-3-030-18472-8}
  {\emph {\bibinfo {title} {Analysis and Data-Based Reconstruction of Complex
  Nonlinear Dynamical Systems}}}\ (\bibinfo  {publisher} {Springer
  International Publishing},\ \bibinfo {year} {2019})\BibitemShut {NoStop}%
\bibitem [{\citenamefont {Applebaum}(2011)}]{Applebaum2011}%
  \BibitemOpen
  \bibfield  {author} {\bibinfo {author} {\bibfnamefont {D.}~\bibnamefont
  {Applebaum}},\ }\href {https://doi.org/10.1017/CBO9780511809781} {\emph
  {\bibinfo {title} {Lévy Processes and Stochastic Calculus}}},\ \bibinfo
  {edition} {2nd}\ ed.\ (\bibinfo  {publisher} {Cambridge University Press},\
  \bibinfo {year} {2011})\BibitemShut {NoStop}%
\bibitem [{EPE(2021)}]{EPEX}%
  \BibitemOpen
  \href@noop {} {\bibinfo {title} {{European Power Exchange (EPEX SPOT)}}}
  (\bibinfo {year} {2021}),\ \bibinfo {note} {{M}arket data
  \href{https://www.epexspot.com/en/market-data}{https://www.epexspot.com/en/market-data}}\BibitemShut
  {NoStop}%
\bibitem [{\citenamefont {{Fraunhofer Institute for Solar Energy System
  ISE}}(2020)}]{PriceData}%
  \BibitemOpen
  \bibfield  {author} {\bibinfo {author} {\bibnamefont {{Fraunhofer Institute
  for Solar Energy System ISE}}},\ }\href@noop {} {\bibinfo {title}
  {{Energy-Charts}}} (\bibinfo {year} {2020}),\ \bibinfo {note}
  {{E}nergy-Charts. 2015--2019
  \href{https://energy-charts.info/charts/price_spot_market/chart.htm}{https://energy-charts.info/charts/price\_spot\_market/chart.htm}}\BibitemShut
  {NoStop}%
\bibitem [{\citenamefont {{SILSO World Data Center, Royal Observatory of
  Belgium, Brussels}}(2020)}]{SunspotData}%
  \BibitemOpen
  \bibfield  {author} {\bibinfo {author} {\bibnamefont {{SILSO World Data
  Center, Royal Observatory of Belgium, Brussels}}},\ }\href@noop {} {\bibinfo
  {title} {{\textit{The International Sunspot Number}}}} (\bibinfo {year}
  {2020}),\ \bibinfo {note} {international Sunspot Number Monthly Bulletin and
  online catalogue: 1818--2020.
  \href{http://www.sidc.be/silso/}{http://www.sidc.be/silso/}}\BibitemShut
  {NoStop}%
\bibitem [{\citenamefont {Stix}(2002)}]{Stix2002}%
  \BibitemOpen
  \bibfield  {author} {\bibinfo {author} {\bibfnamefont {M.}~\bibnamefont
  {Stix}},\ }\href {https://doi.org/10.1007/978-3-642-56042-2} {\emph {\bibinfo
  {title} {The Sun}}},\ \bibinfo {edition} {2nd}\ ed.\ (\bibinfo  {publisher}
  {Springer-Verlag Berlin Heidelberg},\ \bibinfo {year} {2002})\BibitemShut
  {NoStop}%
\bibitem [{\citenamefont {Balogh}\ \emph {et~al.}(2015)\citenamefont {Balogh},
  \citenamefont {Hudson}, \citenamefont {Petrovay},\ and\ \citenamefont {{von
  Steiger}}}]{Andre2015}%
  \BibitemOpen
  \bibinfo {editor} {\bibfnamefont {A.}~\bibnamefont {Balogh}}, \bibinfo
  {editor} {\bibfnamefont {H.}~\bibnamefont {Hudson}}, \bibinfo {editor}
  {\bibfnamefont {K.}~\bibnamefont {Petrovay}},\ and\ \bibinfo {editor}
  {\bibfnamefont {R.}~\bibnamefont {{von Steiger}}},\ eds.,\ \href
  {https://doi.org/10.1007/978-1-4939-2584-1} {\emph {\bibinfo {title} {The
  Solar Activity Cycle}}},\ \bibinfo {edition} {1st}\ ed.\ (\bibinfo
  {publisher} {Springer-Verlag New York},\ \bibinfo {year} {2015})\BibitemShut
  {NoStop}%
\bibitem [{\citenamefont {Braun}\ and\ \citenamefont
  {Brunner}(2018)}]{Braun2018}%
  \BibitemOpen
  \bibfield  {author} {\bibinfo {author} {\bibfnamefont {S.~M.}\ \bibnamefont
  {Braun}}\ and\ \bibinfo {author} {\bibfnamefont {C.}~\bibnamefont
  {Brunner}},\ }\href {https://doi.org/10.1007/s12398-018-0228-0} {\bibfield
  {journal} {\bibinfo  {journal} {Zeitschrift f{\"u}r Energiewirtschaft}\
  }\textbf {\bibinfo {volume} {42}},\ \bibinfo {pages} {257} (\bibinfo {year}
  {2018})}\BibitemShut {NoStop}%
\bibitem [{\citenamefont {Narajewski}\ and\ \citenamefont
  {Ziel}(2020)}]{Narajewski2020}%
  \BibitemOpen
  \bibfield  {author} {\bibinfo {author} {\bibfnamefont {M.}~\bibnamefont
  {Narajewski}}\ and\ \bibinfo {author} {\bibfnamefont {F.}~\bibnamefont
  {Ziel}},\ }\href {https://doi.org/10.1016/j.jcomm.2019.100107} {\bibfield
  {journal} {\bibinfo  {journal} {Journal of Commodity Markets}\ }\textbf
  {\bibinfo {volume} {19}},\ \bibinfo {pages} {100107} (\bibinfo {year}
  {2020})}\BibitemShut {NoStop}%
\bibitem [{\citenamefont {Simonsen}(2003)}]{Simonsen2003}%
  \BibitemOpen
  \bibfield  {author} {\bibinfo {author} {\bibfnamefont {I.}~\bibnamefont
  {Simonsen}},\ }\href {https://doi.org/10.1016/S0378-4371(02)01938-6}
  {\bibfield  {journal} {\bibinfo  {journal} {Physica A: Statistical Mechanics
  and its Applications}\ }\textbf {\bibinfo {volume} {322}},\ \bibinfo {pages}
  {597} (\bibinfo {year} {2003})}\BibitemShut {NoStop}%
\bibitem [{\citenamefont {Weron}\ \emph
  {et~al.}(2004{\natexlab{b}})\citenamefont {Weron}, \citenamefont
  {Bierbrauer},\ and\ \citenamefont {Trück}}]{Weron2004}%
  \BibitemOpen
  \bibfield  {author} {\bibinfo {author} {\bibfnamefont {R.}~\bibnamefont
  {Weron}}, \bibinfo {author} {\bibfnamefont {M.}~\bibnamefont {Bierbrauer}},\
  and\ \bibinfo {author} {\bibfnamefont {S.}~\bibnamefont {Trück}},\ }\href
  {https://doi.org/10.1016/j.physa.2004.01.008} {\bibfield  {journal} {\bibinfo
   {journal} {Physica A: Statistical Mechanics and its Applications}\ }\textbf
  {\bibinfo {volume} {336}},\ \bibinfo {pages} {39} (\bibinfo {year}
  {2004}{\natexlab{b}})},\ \bibinfo {note} {proceedings of the XVIII Max Born
  Symposium ``Statistical Physics outside Physics''}\BibitemShut {NoStop}%
\bibitem [{\citenamefont {Laszuk}(2017)}]{Laszuk2017}%
  \BibitemOpen
  \bibfield  {author} {\bibinfo {author} {\bibfnamefont {D.}~\bibnamefont
  {Laszuk}},\ }\href@noop {} {\bibinfo {title} {Python implementation of
  empirical mode decomposition algorithm}},\ \bibinfo {howpublished}
  {\href{https://github.com/laszukdawid/PyEMD}{https://github.com/laszukdawid/PyEMD}}
  (\bibinfo {year} {2017})\BibitemShut {NoStop}%
\bibitem [{\citenamefont {Hunter}(2007)}]{Matplotlib}%
  \BibitemOpen
  \bibfield  {author} {\bibinfo {author} {\bibfnamefont {J.~D.}\ \bibnamefont
  {Hunter}},\ }\href {https://doi.org/10.1109/MCSE.2007.55} {\bibfield
  {journal} {\bibinfo  {journal} {Computing in Science \& Engineering}\
  }\textbf {\bibinfo {volume} {9}},\ \bibinfo {pages} {90} (\bibinfo {year}
  {2007})}\BibitemShut {NoStop}%
\end{thebibliography}%

\end{document}